\documentclass[twocolumn]{aastex7}
\graphicspath{{./}{figures/}}
\usepackage{amsmath}
\usepackage{subcaption}
\usepackage{gensymb}
\usepackage{xcolor}
\usepackage{soul}
\received{February 10, 2026}
\revised{March 20, 2026}
\accepted{May 11, 2026}
\shorttitle{Deep Learning Photo-$z$'s for Roman}
\shortauthors{Khederlarian et al.}
\begin{document}

\title{Optimizing Deep Learning Photometric Redshifts for the Roman Space Telescope with HST/CANDELS}

\correspondingauthor{Ashod Khederlarian}
\email{ashod\_kh@pitt.edu}

\author[0000-0001-9028-8885]{Ashod Khederlarian}
\affiliation{Department of Physics and Astronomy, University of Pittsburgh, Pittsburgh, PA 15260, USA}
\affiliation{Pittsburgh Particle Physics, Astrophysics, and Cosmology Center (PITT PACC), University of Pittsburgh, Pittsburgh, PA 15260, USA}
\email{ashod\_kh@pitt.edu}

\author[0000-0001-8085-5890]{Brett H.~Andrews}
\affiliation{Department of Physics and Astronomy, University of Pittsburgh, Pittsburgh, PA 15260, USA}
\affiliation{Pittsburgh Particle Physics, Astrophysics, and Cosmology Center (PITT PACC), University of Pittsburgh, Pittsburgh, PA 15260, USA}
\email{andrewsb@pitt.edu}

\author[0000-0001-8684-2222]{Jeffrey A.~Newman}
\affiliation{Department of Physics and Astronomy, University of Pittsburgh, Pittsburgh, PA 15260, USA}
\affiliation{Pittsburgh Particle Physics, Astrophysics, and Cosmology Center (PITT PACC), University of Pittsburgh, Pittsburgh, PA 15260, USA}
\email{janewman@pitt.edu}

\author[0000-0002-5596-198X]{Tianqing Zhang}
\affiliation{Department of Physics and Astronomy, University of Pittsburgh, Pittsburgh, PA 15260, USA}
\affiliation{Pittsburgh Particle Physics, Astrophysics, and Cosmology Center (PITT PACC), University of Pittsburgh, Pittsburgh, PA 15260, USA}
\email{tq.zhang@pitt.edu}

\author[0000-0002-5665-7912]{Biprateep Dey}
\affiliation{Department of Statistical Sciences, University of Toronto, 700 University Ave, Toronto, ON M5G 1Z5, Canada}
\affiliation{Canadian Institute for Theoretical Astrophysics, University of Toronto, 60 St. George Street, Toronto, ON  M5S 3H8}
\affiliation{Dunlap Institute for Astronomy \&\ Astrophysics, University of Toronto, 50 St. George Street, Toronto, ON M5S 3H4, Canada}
\affiliation{Vector Institute for Artificial Intelligence, 108 College St., Toronto, ON M5G 0C6, Canada}
\email{biprateep@pitt.edu}



\begin{abstract}
Photometric redshifts (photo-$z$'s) will be crucial for studies of galaxy evolution, large-scale structure, and transients with the Nancy Grace Roman Space Telescope. Deep learning methods leverage pixel-level information from ground-based images to achieve the best photo-$z$'s for low-redshift galaxies, but their efficacy at higher redshifts with deep, space-based imaging remains largely untested. We used Hubble Space Telescope CANDELS optical and near-infrared imaging to evaluate fully-, self-, and semi-supervised deep learning photo-$z$ algorithms out to $z\sim3$. Compared to template-based and classical machine learning photometry methods, the fully-supervised and semi-supervised models achieved better performance. Our new semi-supervised model, \texttt{PITA} (Photo-$z$ Inference with a Triple-task Algorithm),  outperformed all others by learning from unlabeled and labeled data through a three-part loss function that incorporates images and colors for all objects as well as redshifts when available. \texttt{PITA} produces a latent space that varies smoothly in magnitude, color, and redshift, resulting in the best photo-$z$ performance even when the redshift training set was significantly reduced. In contrast, the self-supervised approach produced a latent space with significant color and redshift fluctuations that hindered photo-$z$ inference. Looking forward to Roman, we recommend using semi-supervised deep learning to take full advantage of the information contained in the hundreds of millions of high-resolution images and color measurements, together with the limited redshift measurements available, to achieve the most accurate photo-$z$ estimates for both faint and bright sources.
\end{abstract}



\section{Introduction} \label{sec:intro}

The Roman Space Telescope \citep{2015arXiv150303757S} is a flagship Stage IV dark energy mission \citep[in the classification system of][]{2006astro.ph..9591A} that will significantly advance our understanding of cosmology and galaxy evolution. With its wide field of view, high survey speed, and high-resolution near-infrared imaging, Roman will collect vast amounts of data that will enable unprecedented studies of weak and strong gravitational lensing, galaxy clustering, galaxy clusters, the evolution of galaxy demographics over time, and transient objects such as type Ia supernovae. A key part of these analyses is knowing the redshifts of detected sources. While reliable redshifts can be measured via spectroscopy (spectroscopic redshifts or spec-$z$'s), this is not possible for most galaxies as spectroscopy is resource-intensive. Therefore, photometric redshift (photo-$z$) measurements will be needed for the vast majority of sources (see \citealt{2019NatAs...3..212S} for a broad review of photo-$z$ methods and \citealt{2022ARA&A..60..363N} for an overview of photo-$z$ needs for next-generation surveys specifically).

Classical photo-$z$ algorithms rely on using photometric colors and magnitudes to estimate the redshift of a galaxy. Template-fitting approaches infer $z$ by fitting observed photometry with a set of redshifted galaxy spectral energy distribution (SED) templates (e.g., LePHARE, \citealt{1999MNRAS.310..540A}, \citealt{2006A&A...457..841I}, \citealt{2009ApJ...690.1236I}, BPZ, \citealt{2000ApJ...536..571B}, ZEBRA \citealt{2006MNRAS.372..565F}, EAZY \citealt{2008ApJ...686.1503B}). While template-based methods offer a physically-motivated means of inferring redshifts and other physical parameters, they are computationally expensive and are limited by the quality of the templates. Synthetic templates rely on model assumptions (e.g., the range of possible star formation histories or dust attenuation laws) and may be limited to too narrow a range of parameter space. On the other hand, empirical templates may lack full wavelength coverage and are difficult to calibrate.

More recent machine learning (ML) methods that empirically learn the relationship between photometry and redshift \citep{2016MNRAS.460.1371B,2016MNRAS.462..726A,2018AJ....155....1G,2021MNRAS.501.3309Z,2021MNRAS.507.5034R,2024AJ....168...80C} offer faster inference and generally outperform template-based approaches \citep{2025arXiv251007370Z}. However, they require large training sets with spec-$z$ labels and are prone to biases in regions of parameter space that are underrepresented or lacking in training sets \citep{2020A&A...644A..31E}. 

Photometric measurements provide physically interpretable, low-dimensional, and high S/N features for estimating galaxy redshifts, but they represent a lossy compression of all the information in an image; any information besides integrated flux (e.g., morphology) is discarded. State-of-the-art photo-$z$ methods instead apply deep neural networks directly to images, leveraging pixel-level information to provide the best photo-$z$'s to date \citep{2019A&A...621A..26P, 2021arXiv210104293A, 2022MNRAS.515.5285D, 2024A&A...692A.260R, 2024A&A...683A..26A, 2024arXiv240901825D,2025arXiv251027032M}. While seemingly promising for Roman, the vast majority of deep learning photo-$z$ algorithms have focused on low-redshift galaxies with ground-based imaging because of the availability of large, fully representative training datasets.

The feasibility of applying deep learning methods to infer photo-$z$'s for the faint, high-redshift galaxies that will dominate Roman samples remains largely untested. This is caused by two main challenges. First, to leverage pixel-level information galaxy morphologies must be well-resolved, but this is not possible for galaxies at redshifts $z > 0.3$ with ground-based imaging due to atmospheric seeing. Second, supervised photo-$z$ algorithms depend on reliable spec-$z$ labels, which are difficult to obtain for fainter and higher-redshift galaxies. 

Roman will address the first of these limitations by imaging thousands of square degrees in multiple bands with high spatial resolution. Additionally, larger samples of reliable redshifts will be available through current and upcoming spectroscopic surveys such as the Dark Energy Spectroscopic Instrument (DESI; \citealt{2025arXiv250314745D}), DESI-II \citep{2022arXiv220903585S}, and the Subaru Prime Focus Spectrograph (PFS; \citealt{2016SPIE.9908E..1MT}), as well as Roman grism and prism observations. However, these redshift labels will mostly be limited to bright sources, such that spectroscopic surveys will lag far behind photometric ones, especially at high redshifts. 

Nonetheless, the sheer volume of Roman's imaging will facilitate training of deep learning photo-$z$ algorithms. Roman's imaging-rich but label-scarce datasets are particularly well suited for self- and semi-supervised algorithms that can leverage unlabeled data, in contrast to fully-supervised methods that only learn from labeled data (i.e., objects with redshift measurements). These methods could potentially learn how to create more informative, less-lossy compressions from the hundreds of millions of Roman images, going beyond the simple information compression provided by catalog-level photometric measurements.

While recent work to apply deep learning methods to Euclid \citep{2025arXiv250315312E} and JWST data \citep{2025arXiv251027032M} is encouraging, both datasets currently have limitations relative to Roman. The Euclid Q1 imaging used to date is much shallower than what Roman will provide, while the spectroscopic labels used in that study are limited to DESI redshifts at $z < 1$ \citep{2024AJ....168...58D}, which is a highly non-representative dataset focused on particular galaxy populations that have easy-to-measure redshifts.  The JWST analysis, in turn, utilized only $\sim2,000$ spec-$z$'s with a highly biased distribution. It therefore remains unclear whether deep learning will yield optimal redshifts for broad galaxy samples from Roman, as the observed galaxy populations will be significantly deeper and higher redshift than those explored for Euclid. The principle aim of this work is to test the efficacy of deep learning photo-$z$ algorithms on Roman-like data and explore potential optimizations. 

In this paper, we use data from the HST/CANDELS imaging survey \citep{2011ApJS..197...35G,2011ApJS..197...36K} as a proxy for Roman. We compare the performance of fully- and self-supervised deep learning algorithms to ML-based or template-based photo-$z$ methods. Additionally, we present our own semi-supervised algorithm \texttt{PITA} (Photo-$z$ Inference with a Triple-task Algorithm), which allows for simultaneous training with unlabeled images, photometric measurements, and redshift labels when available.

% CANDELS is one of the few surveys which has comparable depth, spatial resolution, and wavelength coverage to the medium tier of Roman's high-latitude wide area survey \citep{2025arXiv250510574O}, as well as having a significant subset of galaxies with reliable redshift labels. We used $\sim 100,000$ HST CANDELS galaxies, including $\sim20,000$ with redshift measurements, to compare the performance of different photo-$z$ algorithms.

% We first used photometric measurements to run the template-fitting code LePHARE \citep{2011ascl.soft08009A} and to also train a fully-supervised simple neural network to predict redshift. These acted as baselines to which we compared deep learning methods. As a simple first step, we trained a fully-supervised convolutional neural network (CNN) using only the labeled images to see if pixel-level information can enhance performance. Then, to leverage unlabeled data and potentially obtain even better results, we trained a self-supervised algorithm and fine-tuned it with labels. Finally, we developed our own semi-supervised algorithm \texttt{PITA} (Photo-$z$ Inference with a Triple-loss Algorithm), which allowed us to simultaneously train with unlabeled images, photometric measurements, and redshift labels when available.

The structure of the paper is as follows. In Section \ref{sec:data}, we describe the CANDELS imaging dataset and the redshift labels used in this work. Section \ref{sec:method} describes the methods used for photo-$z$ estimation, including the baseline photometry methods (Section \ref{subsec:photometry_baselines}) and the deep learning algorithms (Sections \ref{subsec:fully_supervised}, \ref{subsec:self_supervised}, and \ref{subsec:pita}). In Section \ref{sec:results}, we compare the photo-$z$ predictions from the different methods and analyze the latent vectors obtained from the self- and semi-supervised algorithms. In Section \ref{sec:discussion}, we discuss the implications of our findings for Roman. Finally, we end with summary and conclusions in Section \ref{sec:summary_and_conclusions}. Throughout this paper, we use AB magnitudes and a flat $\Lambda$CDM cosmology with present-day matter density parameter $\Omega_m=0.3$ and Hubble constant $H_0=70$ km s$^{-1}$ Mpc$^{-1}$.

\section{Data} \label{sec:data}
To test deep learning photo-$z$ algorithms for Roman, we required a Roman-like imaging dataset with reliable redshift labels for a substantial subset of sources. We therefore used the CANDELS imaging survey because it is the only dataset with spatial resolution, depth, and wavelength coverage similar to Roman, while also spanning sufficient sky area to provide enough images and redshift labels to train deep learning algorithms.

\begin{figure*}
    \centering
    \includegraphics[width=1\linewidth]{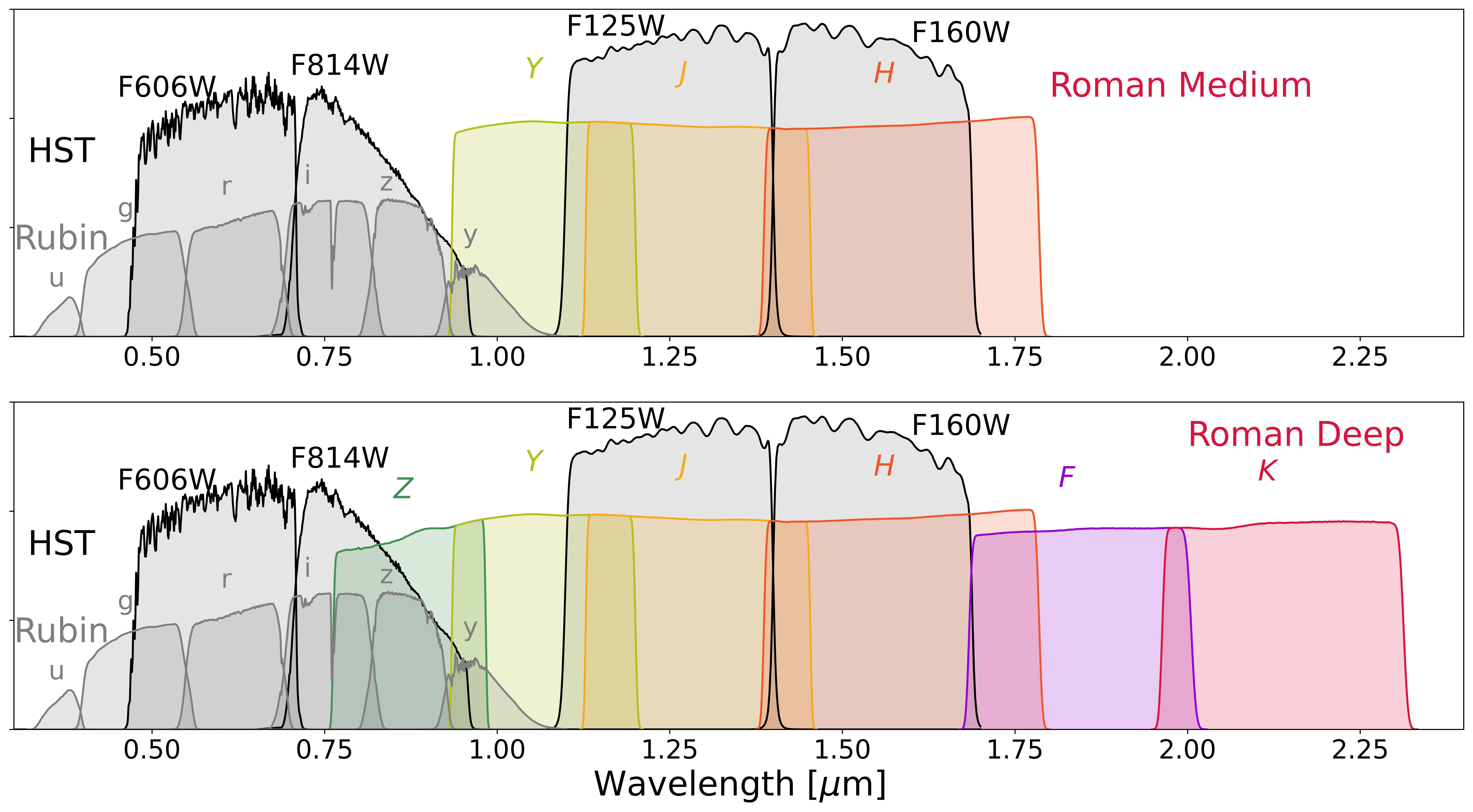}
    \caption{Total system throughputs for HST/CANDELS filters (black), for LSST (gray), and effective areas for Roman $ZYJHFK$ filters (colored). The Roman HLWAS imaging will include all these bands in the deep tier (bottom), but only $YJH$ in the medium tier (top). These bands can be combined with Rubin $ugrizy$ optical photometry to obtain the best possible photo-$z$'s for Roman. In this context, the CANDELS F125W and F160W filters provide near-infrared wavelength coverage comparable to what is available for Roman's medium tier, while F606W and F814 provide optical coverage comparable to the redder Rubin bands, albeit at higher spatial resolution.}
    \label{fig:transmission_curves}
\end{figure*}

\subsection{HST/CANDELS Imaging} \label{subsec:hst_candels_imaging}
CANDELS obtained imaging with HST/ACS in F606W and F814W, and with HST/WFC3 in F125W and F160W, covering a total area of 800 arcmin$^2$ across five sky regions: GOODS-N \citep{2019ApJS..243...22B}, GOODS-S \citep{2013ApJS..207...24G}, UDS \citep{2013ApJS..206...10G}, EGS \citep{2017ApJS..229...32S}, and COSMOS \citep{2017ApJS..228....7N}. Within GOODS-N and GOODS-S, the CANDELS Deep survey spans in total $\sim$125 arcmin$^2$ and reaches a 5$\sigma$ point-source depth of $m_\mathrm{F160W} = 27.7$. The broader CANDELS Wide survey covers all five fields, achieving a 5$\sigma$ point-source depth of $m_\mathrm{F160W} = 27$. 

Our choice of CANDELS was motivated by similarities with Roman's high-latitude wide-area survey (HLWAS), which will provide imaging in four tiers: wide ($2,700$ deg$^2$ in $H$ band only), medium ($2,415$ deg$^2$ in $YJH$), deep ($19.2$ deg$^2$ in $ZYJHFK$), and ultradeep ($5$ deg$^2$ in $YJH$) \citep{2025arXiv250510574O}. The filters $ZYJHFK$ correspond to the official names F087, F106, F129, F158, F184, and F213 respectively. To avoid confusion with HST Filters, we will refer to the Roman filters as $ZYJHFK$ throughout this paper. The medium tier is expected to reach a 5$\sigma$ point-source limit of $m_H=26.5$. The spatial resolution of these images will be $\sim0''.1$ (corresponding to a physical length of $\sim1$kpc at $z=2$), similar to the average full width at half maximum (FWHM) of the HST/ACS point spread functions (PSFs) at the F606W and F814W wavelengths ($0''.115$ and $0''.11$ respectively; \citealt{2011ApJS..197...36K}). 

Figure \ref{fig:transmission_curves} shows the overlap between the effective areas of the Roman $ZYJHFK$ filters and the HST F606W, F814W, F125W, and F160W total system throughputs. We are primarily concerned with photo-$z$'s for the medium tier of HLWAS, for which detected sources are expected to have Rubin \citep{2019ApJ...873..111I} $ugrizy$ optical imaging in addition to Roman $YJH$ near-infrared observations. The CANDELS filters offer a compromise between space-based resolution, depth, and optical-red to near-ir wavelength coverage (which is not the full Roman + Rubin wavelength coverage)

The Multi-band source catalogs from the Mikulski Archive for Space Telescopes (MAST)\footnote{\href{https://archive.stsci.edu/hlsp/candels}{https://archive.stsci.edu/hlsp/candels}} have used \texttt{SExtractor} \citep{1996A&AS..117..393B} for source detection and photometric flux measurement. Across the five CANDELS fields, a total of 186,435 sources were detected in the F160W band. We selected sources that have positive integrated fluxes in all four bands (F606W, F814W, F125W, and F160W) as well as having $m_{\text{F160W}}< 26$, resulting in a catalog of 99,505 sources in total.

To extract image cutouts for these objects, we used the CANDELS image mosaics from MAST, along with the RA and DEC values from the source catalogs. In the ACS bands, which have a pixel resolution of $0''.03$, we extracted $108\times108$ pixel images for each galaxy. In the WFC3 bands, which have a pixel resolution of $0''.06$ (with $\sim0''.199$ average PSF FWHM at both F125W and F160W wavelengths), we extracted $54\times54$ pixels. To place all cutouts on a common pixel grid, each $1 \times 1$ WFC3 pixel was up-sampled to a $2 \times 2$ grid, with the flux distributed evenly among the new pixels. All the images and photometric fluxes were corrected for galactic extinction using dust maps from \cite{2011ApJ...737..103S}, with extinction coefficients $R_\text{F606W}=2.471$, $R_\text{F814W}=1.526$, $R_\text{F125W}=0.726$, and $R_\text{F160W}=0.512$ ($R_V=3.1$).

\begin{figure}
    \centering
    \includegraphics[width=1\linewidth]{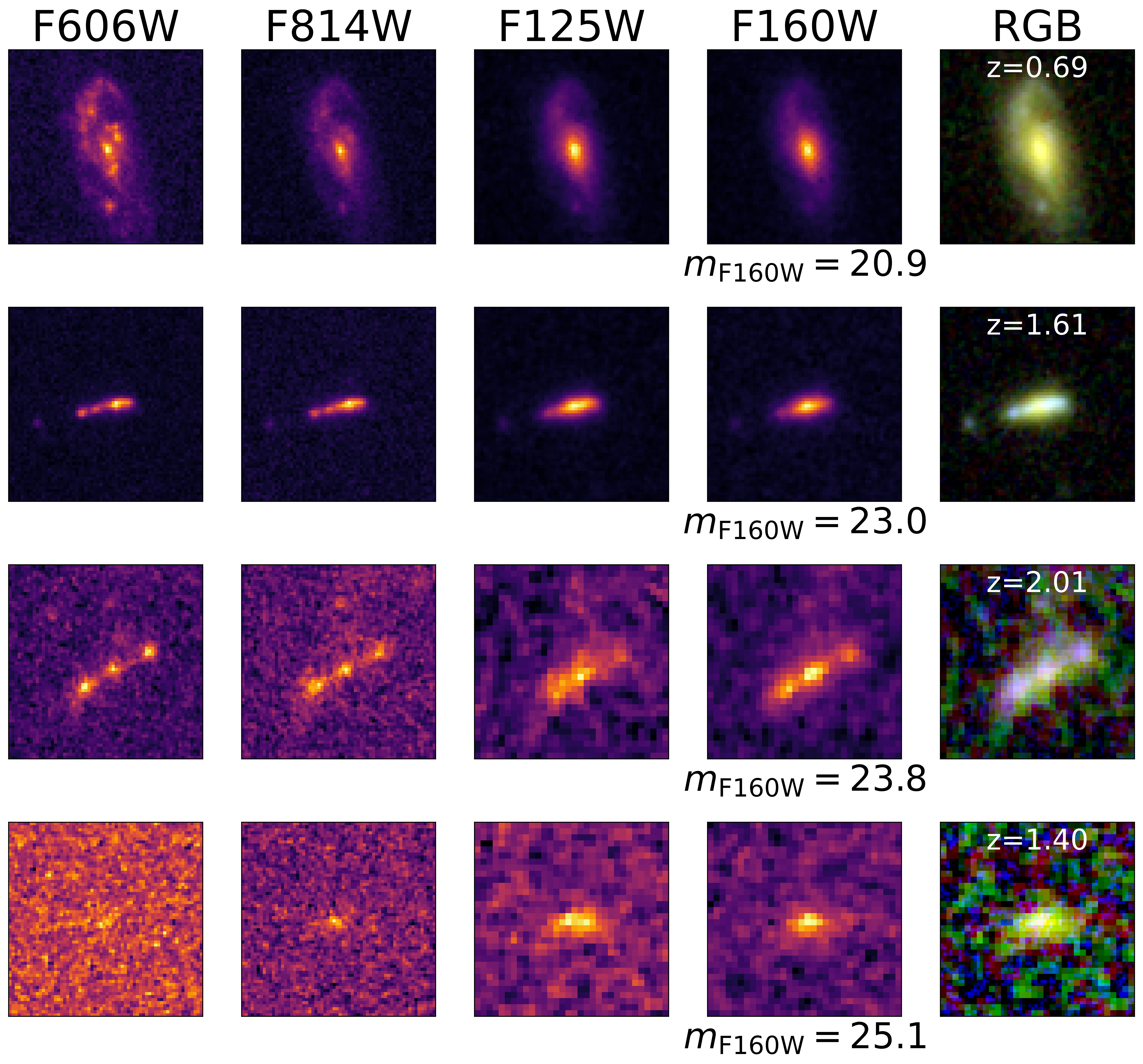}
    \caption{Examples of cutout images for four different galaxies ordered by increasing magnitude in the F160W band. The RGB images are formed using F606W as blue, F125W as green, and F160W as red. For illustrative purposes the pixel values in these images are scaled with an inverse hyperbolic sine function. 
    %We do not apply this scaling during training.
    Galaxies up to $m_\mathrm{F160W}\sim{25}$ have well-measured pixel-to-pixel color variations; this information can potentially be leveraged by deep-learning algorithms.}
    \label{fig:candels_examples}
\end{figure}

Figure \ref{fig:candels_examples} shows example cutout images of objects from CANDELS. The high resolutions of the ACS ($\sim0''.11$ FWHM) and WFC3 ($\sim0''.2$ FWHM) observations provide spatially-resolved images of galaxies with good signal-to-noise ratio (S/N) even for galaxies with faint magnitudes ($m_\mathrm{F160W}\sim25$) and higher redshifts ($z>1$) than those used for previous tests of deep learning photo-$z$'s. The resolution of Roman imaging will be comparable to that achieved by HST at the same wavelength.  

\begin{figure*}
    \centering
    \includegraphics[width=1\linewidth]{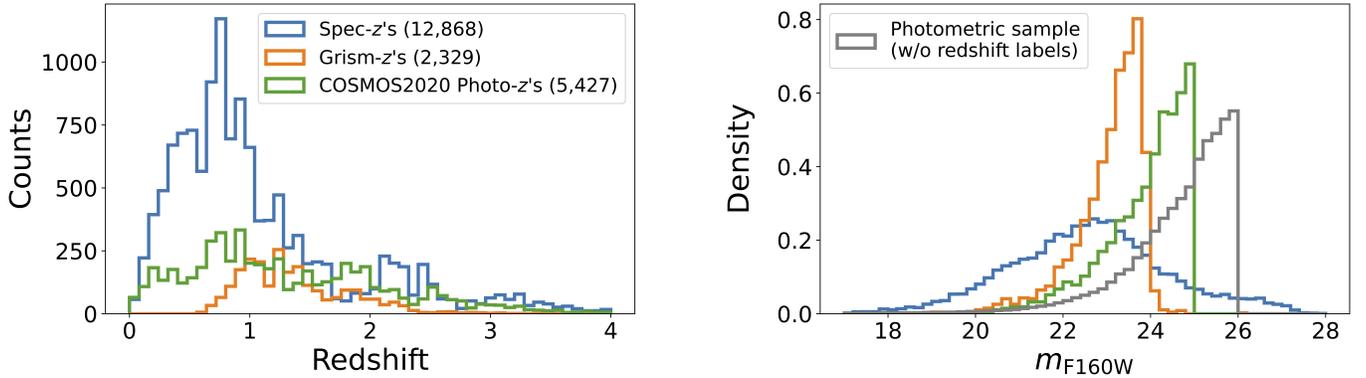}
    \caption{Left: Redshift distributions of the labeled samples (20,624 galaxies) split by spec-$z$'s, grism-$z$'s, and COSMOS2020 many-band photo-$z$'s. Right: $m_\text{F160W}$ distributions of the same labeled samples, in addition to the unlabeled photometric sample ($78,881$ galaxies). The photometric sample is deeper than most of the labeled data.}
    \label{fig:data_distributions}
\end{figure*}

\subsection{Redshift Labels}
The redshift labels we have used in this work are a combination of spec-$z$'s, grism redshifts (grism-$z$'s), and reliable COSMOS2020 many-band photo-$z$'s \citep{2022ApJS..258...11W}. For the 99,505 sources selected from the CANDELS photometric catalogs, we found 9,478 spec-$z$'s across the five CANDELS fields from the compilation made by \cite{2023ApJ...942...36K}, and an additional 3,390 spec-$z$'s in the COSMOS field from the compilation by \cite{2025arXiv250300120K}, bringing the total number of spec-$z$'s to 12,868. We acknowledge the original sources of these spec-$z$'s in Appendix \ref{app:spec_z_citations}. We also used 2,329 grism-$z$'s from \cite{2023ApJ...942...36K} and 5,427 many-band photo-$z$'s from COSMOS2020 (using the \textsc{LePHARE} photo-$z$'s from the classic catalog). The latter were obtained by cross-matching the COSMOS2020 source catalog with the CANDELS-COSMOS source catalog, and keeping only sources with a cross-match separation $\delta<0''.4$ and $m_{\text{F160W}}<25$ (to remove fainter objects which may have unreliable many-band photo-$z$'s). In total, our dataset includes 20,624 redshift labels which we randomly split into training (70\%), validation (15\%), and test (15\%) sets. The images and photometry for the remaining 78,881 unlabeled galaxies were utilized in the self- and semi-supervised training algorithms.

Figure \ref{fig:data_distributions} shows the redshift distributions of the labeled data in the left panel. While objects at lower redshifts ($z<1$) predominate, our dataset includes labels out to $z\sim4$. The right panel shows the $m_\text{F160W}$ distributions of both the labeled and unlabeled data. Generally, the photometric sample is deeper than the labeled data; this will also be the case for Roman. While the brightnesses of the objects with grism-$z$'s or photo-$z$'s are more comparable to the photometric sample, the reliability of their redshift labels gets worse at fainter magnitudes. For this reason, we refer to our redshift labels as $z_\mathrm{ref}$ (short for $z_\mathrm{reference}$) instead of $z_\mathrm{true}$. Spec-$z$'s are more reliable; however, their coverage is biased towards brighter and lower-redshift galaxies \citep{2025arXiv251209032A}.

\section{Methods} \label{sec:method}
The aim of this work is to use the Roman-like HST CANDELS dataset to investigate whether deep learning applied directly on space-based images can yield better photo-$z$ predictions than photometry-based methods using the same data. While this has been found to be true for very low-redshift galaxies with ground-based imaging, our dataset comprises significantly different populations of galaxies that have a wider range of magnitudes, morphologies, and redshifts. Therefore, it is not clear \textit{a priori} whether deep learning will offer improvements in this deeper, higher-redshift regime.

To address this question, we used classical ML and template-fitting predictions from photometry alone as baselines against which we compared three different deep learning algorithms. First, we trained a fully-supervised CNN to predict redshift from four-band image inputs. This is the simplest deep learning approach that relies solely on labeled data. Then, we explored pre-training with unlabeled data and fine-tuning with redshift labels. Finally, we developed our own semi-supervised algorithm, \texttt{PITA}, that is simultaneously trained on both unlabeled and labeled data. All machine learning methods were implemented in Python with \texttt{PyTorch} \citep{2019arXiv191201703P}. To train the deep learning algorithms on multiple GPUs, we used \texttt{PyTorch Lightning} \citep{Falcon_PyTorch_Lightning_2019}.

\subsection{Classical (Photometry-only) Methods} \label{subsec:photometry_baselines}
\subsubsection{Simple Neural Network} \label{subsubsec:ML_baseline}
For our first baseline, we trained a fully-supervised ML photo-$z$ method using only the labeled data to learn a mapping between integrated photometry and redshift. Specifically, we employed a multi-layered perceptron (MLP), which is a simple neural network consisting of seven hidden layers with [128, 128, 128, 128, 128, 64, 32] neurons (resulting in $77,057$ parameters). The network takes as input a four-dimensional vector composed of three colors (F606W$-$F814W, F814W$-$F125W, F125W$-$F160W) and a magnitude ($m_\mathrm{F160W}$). The output is a single redshift value. 

To train the network, we used a Huber loss function \citep{huber1992robust}, which behaves as an L2 loss for prediction errors less than or equal to the value of the transition parameter ($\delta$) and converges to an L1 loss for prediction errors greater than $\delta$:

\begin{equation} \label{eq:redshift_loss}
L_\delta(y,f(\boldsymbol{x}))=\begin{cases}
\frac{1}{2}(y-f(\boldsymbol{x}))^2, & \text{if $|y-f(\boldsymbol{x})|\leq\delta$},\\
\delta(|y-f(\boldsymbol{x})|-\frac{1}{2}\delta), & \text{if $|y-f(\boldsymbol{x})|>\delta$},
\end{cases}
\end{equation}

\noindent where $y$ denotes the true redshift and $f(\boldsymbol{x})$ denotes the predicted redshift for input vector $\boldsymbol{x}$. We set the transition parameter to $\delta=0.15$ (motivated by the catastrophic outlier cutoff being $\frac{|z_\mathrm{ref}-z_\mathrm{pred}|}{1+z_\mathrm{ref}} > 0.15$).

To minimize the loss function, we used the Adam optimizer \citep{kingma2017adammethodstochasticoptimization} with no weight decay, a learning rate of $10^{-3}$, and a batch size of $512$. We trained an ensemble of 10 networks for 200 epochs each, and used the network and epoch that had the minimum redshift validation loss to calculate photo-$z$'s. 

\subsubsection{LePHARE Template Fitting} \label{subsubsec:template_baseline}
As a template-fitting baseline, we used the implementation of \textsc{LePhare} \citep{Arnouts1999,Ilbert2006} in the Redshift Infrastructure Assessment Layers (RAIL; \citealt{rail2025}) package\footnote{\url{https://github.com/LSSTDESC/rail_lephare}} to estimate photometric redshifts from the same input photometry as was used for our simple neural network. \textsc{LePhare} computes model fluxes by redshifting a library of SED templates, applying dust attenuation and emission-line prescriptions, and fitting each template to the observed multi-band photometry. At each redshift, the minimum $\chi^2$ over all tempaltes is used to construct a likelihood function, effectively forming a profile likelihood over tempaltes evaluated on the redshift grid. The posterior redshift probability distribution $P(z)$ is then constructed within a Bayesian framework by multiplying the estimated likelihood by a magnitude-dependent prior.

More specifically, we applied \textsc{LePhare} to the AB magnitude from the four CANDELS bands (F606W, F814W, F125W, and F160W), using the COSMOS galaxy template set \citep{Ilbert2009}. We enabled the zero-point adaptation option in \textsc{LePhare}, allowing small magnitude offsets calculated via the residuals of model magnitudes from the observed magnitudes of the training set. A magnitude-dependent Bayesian prior was applied using the F814W band as the reference magnitude. Redshifts were fit over the range $0 \le z \le 6$ with a grid of $\Delta z = 0.01$.

\subsection{Fully-supervised CNN} \label{subsec:fully_supervised}
% \begin{figure*}
%     \centering
%     \includegraphics[width=0.95\linewidth]{fully_supervised_arch_new.png}
%     \caption{Architecture of the fully-supervised redshift prediction model. A ConvNeXt CNN takes as input a four-band image and outputs a 1000-dimensional feature vector, which is then passed to an MLP that predicts a scalar redshift. This network is trained exclusively on galaxies with redshift labels.}
%     \label{fig:fully_supervised_arch}
% \end{figure*}

As the first deep learning algorithm, we trained a ConvNeXt CNN \citep{liu2022convnet} in a fully-supervised fashion to predict photo-$z$'s directly from images. While vision transformers (ViTs; \citealt{dosovitskiy2020image}) can outperform CNNs in some contexts, they are data-hungry and not optimal for our limited dataset size. ConvNeXt is a deep CNN that is based on the ResNet design \citep{he2016deep}, but modernized to incorporate improvements from ViTs, making it ideal for our dataset. 
Specifically, we used the ``tiny" model of ConvNeXt as implemented in PyTorch.\footnote{\url{https://pytorch.org/vision/main/models/convnext.html}} This model outputs a 1,000-dimensional vector which we projected onto a scalar redshift using an MLP consisting of seven hidden layers with [512, 128, 128, 128, 128, 128, 64] neurons. In total, this network has $\sim30$ million parameters.

\begin{figure*}
    \centering
    \includegraphics[width=1\linewidth]{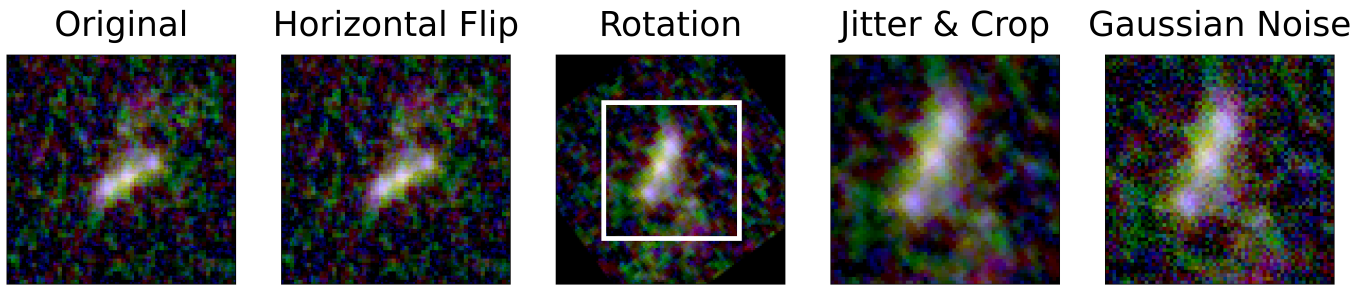}
    \caption{Illustration of the data augmentations used when training the fully-supervised algorithm. Starting with a $108\times108$ pixel image, we apply a horizontal flip with 50\% probability, followed by a random rotation. Then, we jitter the image and crop to the central $64\times64$ pixels. Lastly, we add uncorrelated Gaussian noise. This figure illustrates the effects of the transformations on an RGB composite image, but during training the same transformations are applied on each of the four bands. These transformations make the network more robust and generalizable, and importantly they do not change the characteristics of the images which should provide information about redshift  (e.g., color and morphology).}
    \label{fig:separate_transforms}
\end{figure*}

To help the network generalize and obtain robust results, we applied a set of random augmentations to the images during training: specifically, flips, rotations, jitter and crop, and Gaussian noise. We began with $108\times108$ pixel images and applied horizontal flips with 50\% probability, in addition to random rotations with bilinear interpolation. We also applied random shifts to the image center (jitter) by sampling two integers from a uniform distribution over the range $(-4,4)$ pixels along each axis. We then cropped the image to $64\times64$ pixels and finally added uncorrelated Gaussian noise to each pixel drawn from a distribution with mean zero and band-dependent standard deviation $\sigma_{\text{band}}$. The value of $\sigma_{\text{band}}$ was set to the median absolute deviation of the fluxes in a given band across all pixels and all training data.

Figure \ref{fig:separate_transforms} illustrates the effects of the sequential application of these transformations to an RGB galaxy image. When training, the same transformations are applied to all four bands of a $4\times108\times108$ pixel image tensor, resulting in a $4\times64\times64$ image that serves as the input.

To train the network, we again used the Huber loss function for redshift prediction given by Eq.~\ref{eq:redshift_loss} with $\delta=0.15$; the Adam optimizer with a weight decay of $10^{-5}$; a batch size of 128; and a learning rate that linearly rises from $10^{-5}$ to $5\times10^{-4}$ within 100 epochs, decays as a cosine to $5\times10^{-6}$ across 500 epochs (which is half the period), and then remains constant. We trained 20 networks for 1,000 epochs and chose the network and epoch (selected from the last 100 for each) that had the lowest redshift validation loss.

% Convolutional Neural Networks (CNNs) offer a promising way to extract from images pixel-level information that is relevant to the task at hand \citep{simonyan2014very}. Applications in astronomy for redshift prediction have mostly relied on using ResNet-like architectures that were highly successful in the 2010s (AlexNet \citealt{krizhevsky2012imagenet}, ResNet \citealt{he2016deep}, DesneNet \citealt{huang2017densely}). In particular, ResNet is based on convolutional building blocks that include weight layers, ReLU activation functions \citep{glorot2011deep}, Batch Normalization layers \citep{ioffe2015batch}, and residual connections. Many such blocks are connected, sometimes with average and max pooling layers, to form deep convolutional neural networks with millions of parameters (See \cite{prince2023understanding} for an overview of deep learning). 

% The success of the Transformer architecture \citep{46201} in natural language processing prompted the adaptation to computer vision tasks via Vision Transformers \citep{dosovitskiy2020image}. While ViTs outperform traditional CNNs, they require significantly larger amounts of training data. Since we have limited space-based imaging with reliable redshift measurements, we have not adopted ViTs for this project. In the future, with Roman and Euclid imaging, ViTs can potentially be applied in astronomy. Therefore, we stick with CNN architectures for their data efficiency (because sliding windows that better capture local features, and translational equivariance).

\subsection{Self-supervised CNN} \label{subsec:self_supervised}

Both the ML-photometry baseline and the fully-supervised CNN described above were trained solely on labeled data (i.e., objects with measured redshifts). To take advantage of the significantly larger set of unlabeled data available, we adopted a self-supervised approach inspired by the work of \cite{2021arXiv210104293A}, which demonstrated that photo-$z$ predictions can be improved beyond fully-supervised deep learning methods by first training on all available images in a self-supervised manner, and then fine-tuning for redshift prediction using a labeled subset.

Self-supervised learning is a strategy for machine learning which can make use of unlabeled data. Unlike fully-supervised learning, where the loss function directly compares predicted labels to ground-truth values, self-supervised techniques create pseudo-labels from the data or rely on the training of surrogate tasks. Common examples of such tasks include image reconstruction, predicting masked pixels, or creating different images of the same object and identifying them as a pair (see \citealt{2023arXiv230105712G} for a review of self-supervised learning). In most cases, the aim of a self-supervised strategy is to learn latent representations of the data that are useful for downstream tasks such as classification or regression. While our main concern in this paper is photo-$z$ prediction, general-purpose self-supervised models (known as foundation models; see \citealt{2021arXiv210807258B}) offer a promising way of analyzing large-scale multi-modal astronomical data. We discuss the results of this work within the context of foundation models in Section \ref{sec:discussion}.

\subsubsection{Contrastive Learning}

A particularly promising self-supervised technique is contrastive learning, which has been successfully applied in astronomy previously (see \citealt{2023RASTI...2..441H} for a review). This method has proven effective in learning latent vectors of multi-band images that are informative for redshift prediction in other works \citep{2021arXiv210104293A,2024MNRAS.531.4990P}. Building on those successes, which focused on low-redshift galaxies with ground-based imaging, we implemented contrastive learning on our CANDELS dataset to assess its potential in higher-redshift regimes with space-based imaging.

Contrastive learning offers a framework for constructing latent vectors of galaxy images that are invariant to specific transformations or views of the data. The goal is to learn representations that capture intrinsic features of galaxies while remaining robust to augmentations such as flips, crops, or noise.

In practice, our implementation followed that of MoCo \citep{he2020momentum}. We started by augmenting each image $x$ using a chosen set of transformations ($T$) to generate new images $x^t = T(x)$. Augmented images derived from the same initial image ($x^t_i, x^t_{i^+} = T(x_i), T(x_i)$) form a ``positive pair'', while those originating from different initial images ($x^t_i, x^t_j = T(x_i), T(x_j)$) are labeled as ``negative pairs''. The training process then aims to maximize some measure of similarity between positive pairs in latent space. Typically, the cosine similarity ($S_C$), defined between two vectors ($\boldsymbol{A}$, $\boldsymbol{B}$), is used:
\begin{equation}
    S_C(\boldsymbol{A},\boldsymbol{B}) = \frac{\boldsymbol{A}\cdot\boldsymbol{B}}{||\boldsymbol{A}||\;||\boldsymbol{B}||}, 
\end{equation}
where $||\boldsymbol{A}||=\sum_i{A_i}^2$ is the magnitude of $\boldsymbol{A}$. To prevent the trivial solution of collapsing all images to the same latent vector, the loss function incorporates an additional repulsive term that minimizes the similarity between negative pairs.

In MoCo, contrastive learning is framed as a dictionary look-up task. Given the latent vector $\boldsymbol{q}$ of an image (obtained by applying the encoder on the transformed image), the latent vector $\boldsymbol{k_+}$ of its positive pair, and the latent vectors $\boldsymbol{k_i}$ of the remaining negative-pair images in the dictionary ($K$ images), the InfoNCE loss function \citep{oord2018representation} used is defined as: 

\begin{equation} \label{eq:contrastive_loss}
    l_{q} = -\text{log}\frac{\exp(S_C(\boldsymbol{q},\boldsymbol{k}_+)/\tau)}
    {\sum_{i=0}^{K}\exp(S_C(\boldsymbol{q},\boldsymbol{k}_i)/\tau)}.
\end{equation}

\noindent Here, $\tau$ is a temperature hyperparameter, and the sum is performed over all the vectors in the dictionary, including the positive one $\boldsymbol{k_+}$. This loss function can be interpreted as the categorical cross-entropy loss function for classifying $\boldsymbol{q}$ as being the positive pair of $\boldsymbol{k_+}$. The total loss for a training batch is computed as the average of the losses $\boldsymbol{l_q}$ for each query image. The latent vectors of the transformed images were obtained using a ConvNeXt CNN encoder. Additional details of training this algorithm can be found in Appendix \ref{app:contrastive_learning}.

Figure \ref{fig:candels_view1_view2} illustrates how the transformations described in section \ref{subsec:fully_supervised} are used to construct multiple views of the same galaxy, which serve as the positive pairs. Choosing the correct set of transformations is crucial for obtaining a latent space appropriate for the downstream photo-$z$ prediction task; it is important to avoid any augmentations that can potentially alter redshift information.

\begin{figure}
    \centering
    \includegraphics[width=1\linewidth]{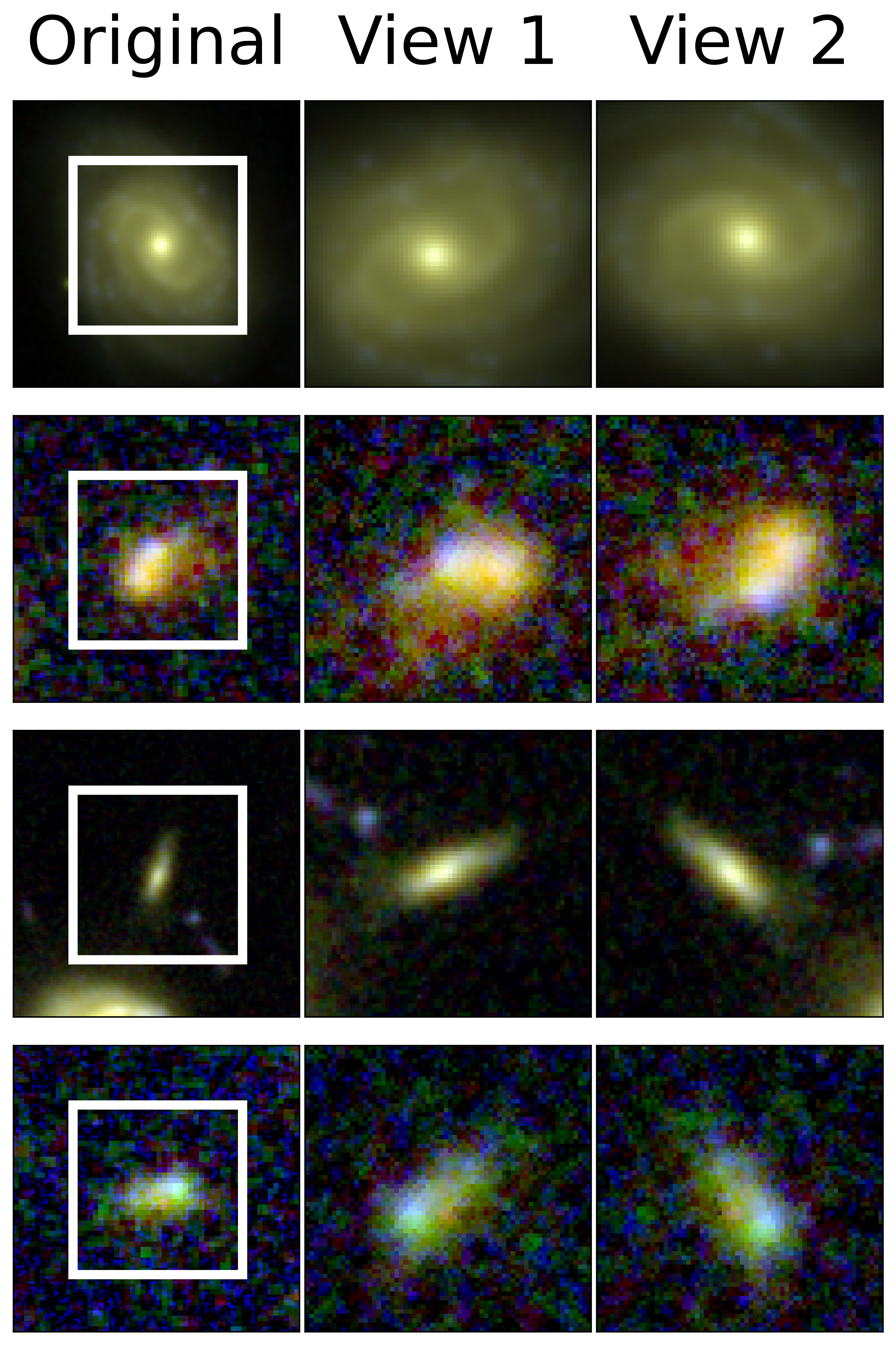}
    \caption{The random transformations illustrated in Figure \ref{fig:separate_transforms} are applied to the original galaxy images to generate two distinct positive views of each galaxy. This figure illustrates these transformations using RGB composite images, but during training the transformations are consistently applied to each of the four bands. Since contrastive learning aims to maximize the similarity between two views, we did not include any transformations that can potentially alter redshift information, such as  augmentations that change colors.}
    \label{fig:candels_view1_view2}
\end{figure}

\begin{figure*}
    \centering
    \includegraphics[width=1\linewidth]{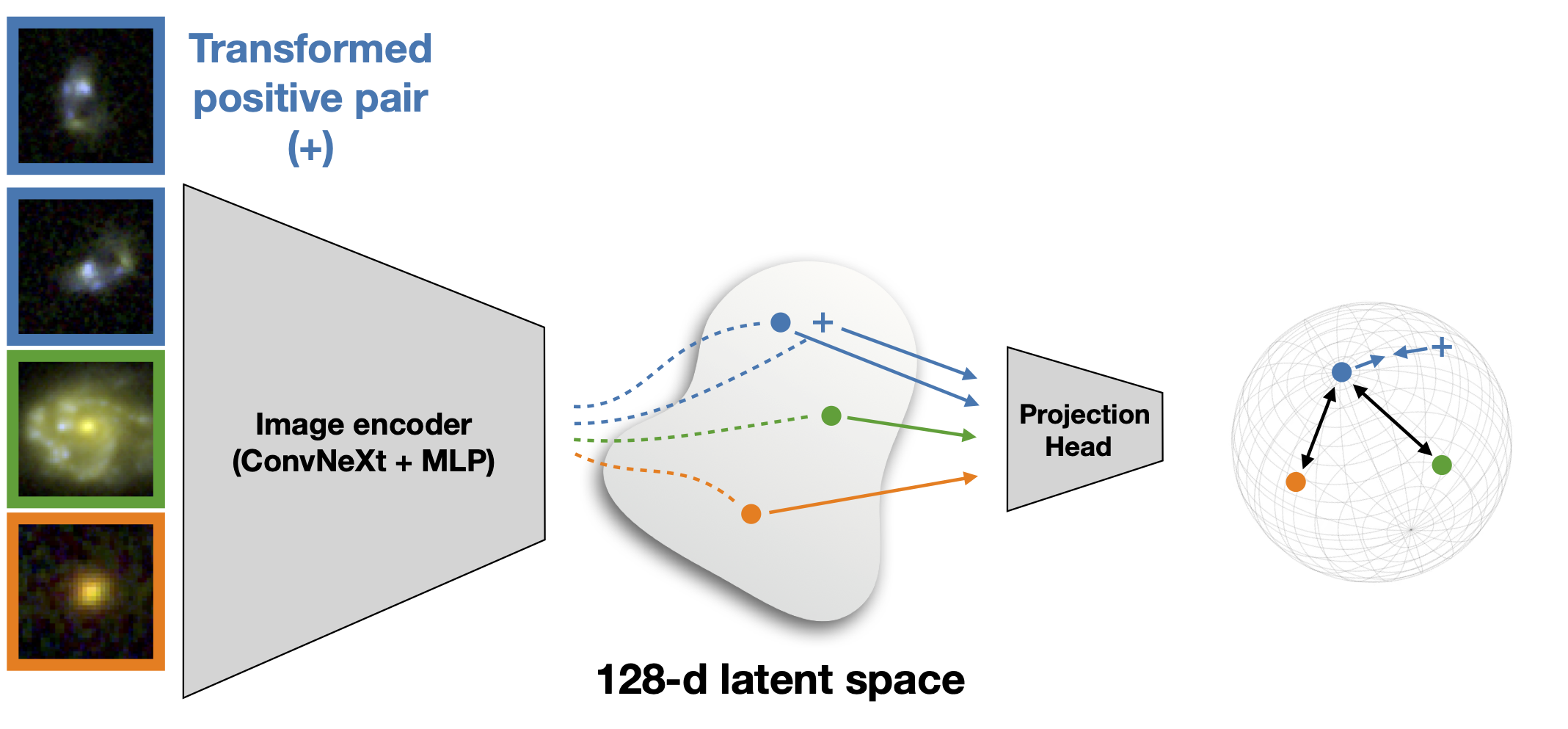}
    \caption{Architecture of the self-supervised contrastive learning algorithm. An image encoder, consisting of a ConvNeXt CNN followed by an MLP, maps four-band galaxy images into a 128-dimensional latent space. A projection head then maps these latent vectors to a 64-dimensional space where the contrastive loss is calculated. This network is initially trained on unlabeled data with the objective of learning image representations that are similar for positive pairs and dissimilar for negative pairs. After this pre-training phase, the projection head is replaced with another $\mathrm{MLP}$, and the whole network is fine-tuned to predict redshifts using only the labeled data.}
    \label{fig:self_supervised_arch}
\end{figure*}

Figure \ref{fig:self_supervised_arch} shows the self-supervised model architecture employed here. We used a second network, which we label as $\mathrm{MLP}_\mathrm{encoder}$ to project the 1000-dimensional output of ConvNeXt into a 128-dimensional latent space. Another MLP, referred to as the projection head, then maps the latent vectors to a final 64-dimensional space. While the contrastive loss is calculated in this projected space, prior studies have found that representations in the latent space (before the projection head) tend to perform better on downstream tasks \citep{chen2020simple,he2020momentum}; we found similar behavior in the process of this work.

To train the network, we used the Adam optimizer with a weight decay of $10^{-5}$; a batch size of 128; a dictionary size of $K=60,000$; and a learning rate that linearly warms up from $10^{-5}$ to $10^{-3}$ within 200 epochs, decays as a cosine back to $10^{-5}$ within 500 epochs (which is half the period), and then stays constant. We trained the network for 9,000 epochs. While the loss function slowly kept decreasing when training longer, we found no noticeable improvement in downstream photo-$z$ prediction performance. 

\subsubsection{Fine-tuning for Redshift Prediction}
To fine-tune the network for redshift prediction, we replaced the projection head with a separate MLP, which we label $\mathrm{MLP}_\mathrm{redshift}$, that maps the 128-dimensional latent vectors to a scalar redshift value. This redshift network consists of six hidden layers with [128, 128, 128, 128, 64, 32] neurons, and is initialized with random weights. The ConvNeXt CNN and $\text{MLP}_\text{encoder}$ are initialized with the weights learned during the self-supervised pretraining stage.

The network is then trained with the $\sim14,000$ labeled images in our training set, using the Adam optimizer with a weight decay of $10^{-5}$, a batch size of 128, and the same learning rate scheduler as in the fully-supervised case. We also applied the same set of transformations described in Section \ref{subsec:fully_supervised}. In effect, this fine-tuning step is the same as training a fully-supervised algorithm with the encoder weights all initialized to the values learned from the contrastive pre-training.

While self-supervised pretraining followed by fine-tuning has been found to be effective at lower redshifts, it was not clear \textit{a priori} that the same approach would work for our dataset; if the contrastive learning objective does not help to distinguish amongst different potential high redshift values, subsequent fine-tuning will not be enough to align the latent space. Indeed, the performance of this algorithm was worse than the fully-supervised CNN. We elaborate on this result and discuss its implications in Sections \ref{sec:results} \& \ref{sec:discussion}. 

\subsection{Semi-supervised CNN: \texttt{PITA}} \label{subsec:pita}
\begin{figure*}
    \centering
    \includegraphics[width=0.95\linewidth]{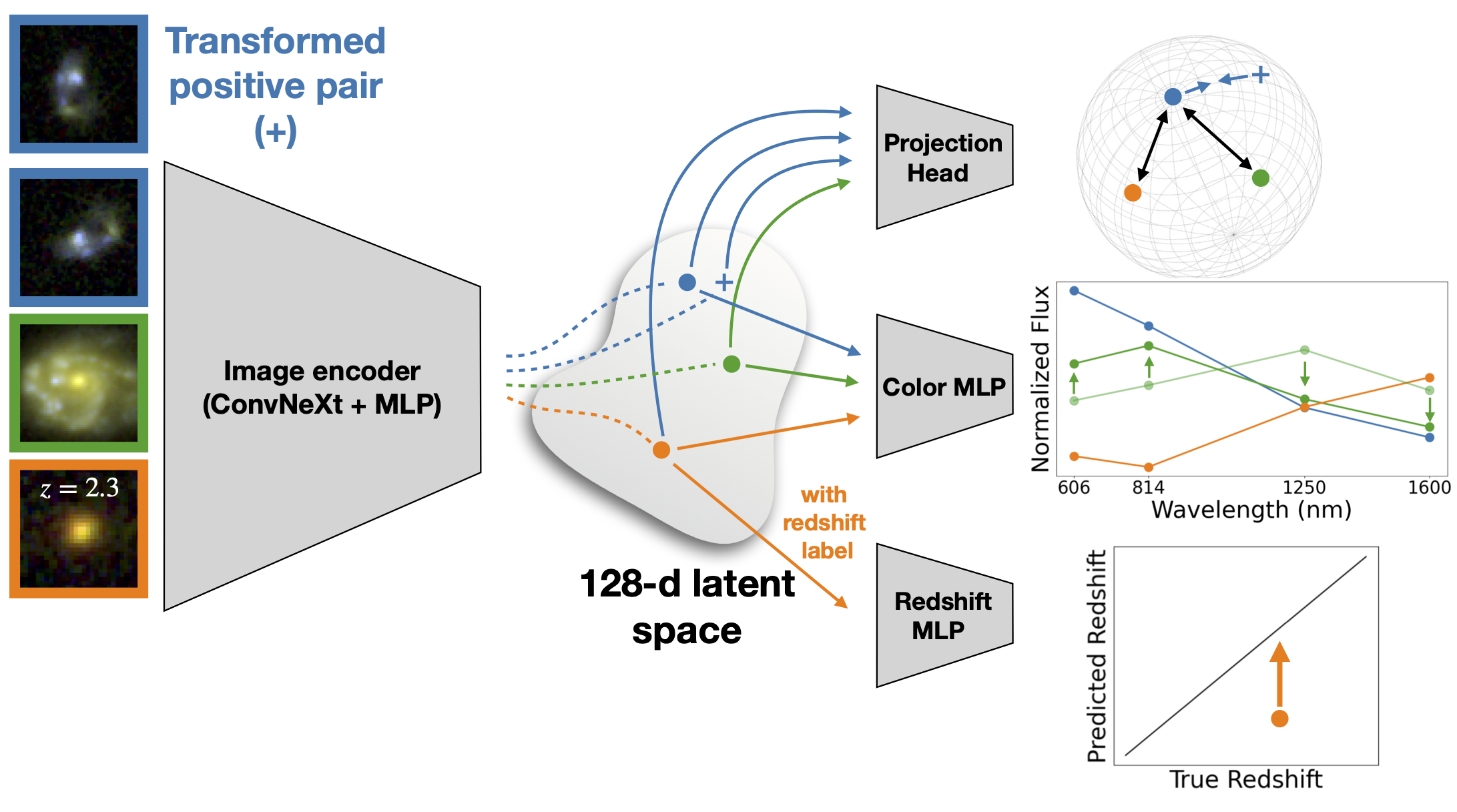}
    \caption{Architecture of \texttt{PITA}. Input images are processed by an encoder consisting of a ConvNeXt CNN followed by an MLP to obtain 128-dimensional latent vectors. Three separate task-specific MLPs further project these representations: (1) a projection head which reduces the dimensionality for contrastive learning; (2) a color MLP used to predict photometric colors for all objects; and (3) a redshift MLP for predicting redshift, which is applied only to those galaxies with labeled redshifts. By combining the loss functions from the three MLPs, both labeled and unlabeled images can be used to learn a latent space that encodes morphology, color, and redshift information.}
    \label{fig:semi_supervised_arch}
\end{figure*}

In analyzing the outputs of the self-supervised model introduced in Section \ref{subsec:self_supervised}, we found that the latent vectors it generated did not capture color information well; however, galaxy colors are highly sensitive to and informative about redshift, implying that the latent vectors were poorly encoding key information. Although subsequent fine-tuning using the labeled data improved the representation of colors to some extent, overall performance remained suboptimal. This motivated us to develop the semi-supervised algorithm \texttt{PITA}.

Semi-supervised learning can combine the best aspects of self- and fully-supervised approaches. In our application, a self-supervised contrastive loss is used by leveraging all available images to obtain latent vectors for galaxies. Simultaneously, we can use the color of all the galaxies and the redshifts for the much smaller labeled subset to align these latent vectors, making them more informative for predicting photo-$z$'s.

\texttt{PITA} jointly optimizes a contrastive loss $L_{\text{contrastive}}$, a color prediction loss $L_{\text{color}}$, and a redshift prediction loss $L_{\text{z}}$ to yield a combined loss, $L_\mathrm{\texttt{PITA}}$:

\begin{equation}
    L_\mathrm{\texttt{PITA}} = L_{\text{contrastive}} + L_{\text{color}} + L_{\text{z}}.
\end{equation}

\noindent The three losses are normalized such that they have the same order of magnitude despite the redshift loss being calculated for a much smaller sample. We discuss the impact of different loss-function weighting schemes in Appendix \ref{app:loss_weights}.

The contrastive loss $L_{\text{contrastive}}$ is the average of Eq.~\ref{eq:contrastive_loss} over all images in the batch. For the redshift loss, we used the Huber loss function introduced in Eq.~\ref{eq:redshift_loss} with $\delta=0.15$. For the color loss we use an L2 loss function,
\begin{equation}
    L_{\text{color}} = \frac{1}{N}\sum_{i=1}^{N}|\boldsymbol{y}_i-\boldsymbol{f}(\boldsymbol{x}_i)|^2,
\end{equation}
where $N$ is the batch size, $\boldsymbol{y}_i$ is the true value, and $\boldsymbol{f}(\boldsymbol{x}_i)$ is the predicted value. Here, $\boldsymbol{y}_i$ is a 4-d vector representing the three colors (F606W$-$F814W, F814W$-$F125W, and F125W$-$F160W), and a single magnitude ($m_\text{F160W}$). 

The contrastive loss does not require any labels to evaluate.  Similarly, colors can be calculated for every object, so all the data within a batch contribute to these two loss terms. However, only images with labels contribute to the redshift loss. If there are no labeled images within a batch, the redshift loss is set to zero.

\subsubsection{\texttt{PITA} Architecture}
Figure \ref{fig:semi_supervised_arch} illustrates the architecture used for \texttt{PITA}. Similar to the previous architectures, the encoder consists of a ConvNext CNN followed by an $\text{MLP}_\text{encoder}$, which together process the four-band input images and output 128-dimensional latent vectors. From this shared latent space, three task-specific MLPs further project these representations to calculate the loss: 

\begin{itemize}
    \item the \textbf{projection head}, which is a single-hidden-layer MLP with 128 neurons that maps the 128-dimensional latent vector to a 64-dimensional space where the contrastive loss is computed;
    \item \textbf{$\text{MLP}_\text{color}$,} which consists of six hidden layers with [128, 128, 128, 128, 128, 64] neurons and outputs a 4-dimensional vector corresponding to F606W$-$F814W, F814W$-$F125W, F125W$-$F160W, and $m_\text{F160W}$; and
    \item \textbf{$\text{MLP}_\text{redshift}$,} which also consists of six hidden layers with [128, 128, 128, 128, 128, 64] neurons and outputs a scalar redshift.
\end{itemize}

To train \texttt{PITA}, we used the Adam optimizer with a weight decay of $10^{-5}$; a batch size of 128; a dictionary size of 60,000; and an oscillating learning rate from $10^{-3}$ (starting value) to $10^{-5}$ with a period of 1000 epochs. Since contrastive learning algorithms are susceptible to local minima and feature suppression \citep{chen2021intriguing,li2023addressing}, we trained 20 networks for 1,500 epochs and chose the single network and epoch that had the minimum redshift validation loss across the last 100 epochs of each.

\section{Results} \label{sec:results}

\begin{deluxetable*}{lcccccc}
    \tablecaption{Photo-$z$ Performance\label{tab:metrics_table}}
    \tablehead{\colhead{} & \multicolumn{3}{c}{$m_{\text{F160W}}<22$} & \multicolumn{3}{c}{$m_{\text{F160W}}<25$} \\
               \multicolumn{1}{l}{Method} & \colhead{bias} & \colhead{$\sigma_{\text{NMAD}}$} & \colhead{$f_{\text{outlier}}$} & \colhead{bias} & \colhead{$\sigma_{\text{NMAD}}$} & \colhead{$f_{\text{outlier}}$}
    }
    \startdata
    %\hline
    LePHARE (photometry) & -0.024 & 0.065 & 10.4\% & -0.006 & 0.082 & 22.3\% \\
    %\hline
    MLP (photometry) & 0.006 & 0.040 & 6.9\% & 0.035 & 0.059 & 14.8\% \\
    %\hline
    Fully-supervised & 0.006 & 0.037 & 3.9\% & 0.026 & 0.056 & \textbf{12.3\%} \\
    %\hline
    Self-supervised & 0.014 & 0.046 & 5.9\% & 0.036 & 0.066 & 14.6\% \\
    %\hline
    \texttt{PITA} (semi-supervised) & \textbf{-0.003} & \textbf{0.033} & \textbf{3.0\%} & \textbf{0.023} & \textbf{0.051} & \textbf{12.3\%} \\
    %\hline
    \enddata
    \tablecomments{Photo-$z$ performance metrics (bias, $\sigma_{\text{NMAD}}$, and $f_{\text{outlier}}$) for each method calculated using test set sources brighter than {$m_{\text{F160W}}=22$} (left column) or brighter than {$m_{\text{F160W}}=25$} (right column). The best values in each column are highlighted in bold.}
\end{deluxetable*}

\begin{figure*}
    \centering
    \includegraphics[width=0.8\linewidth]{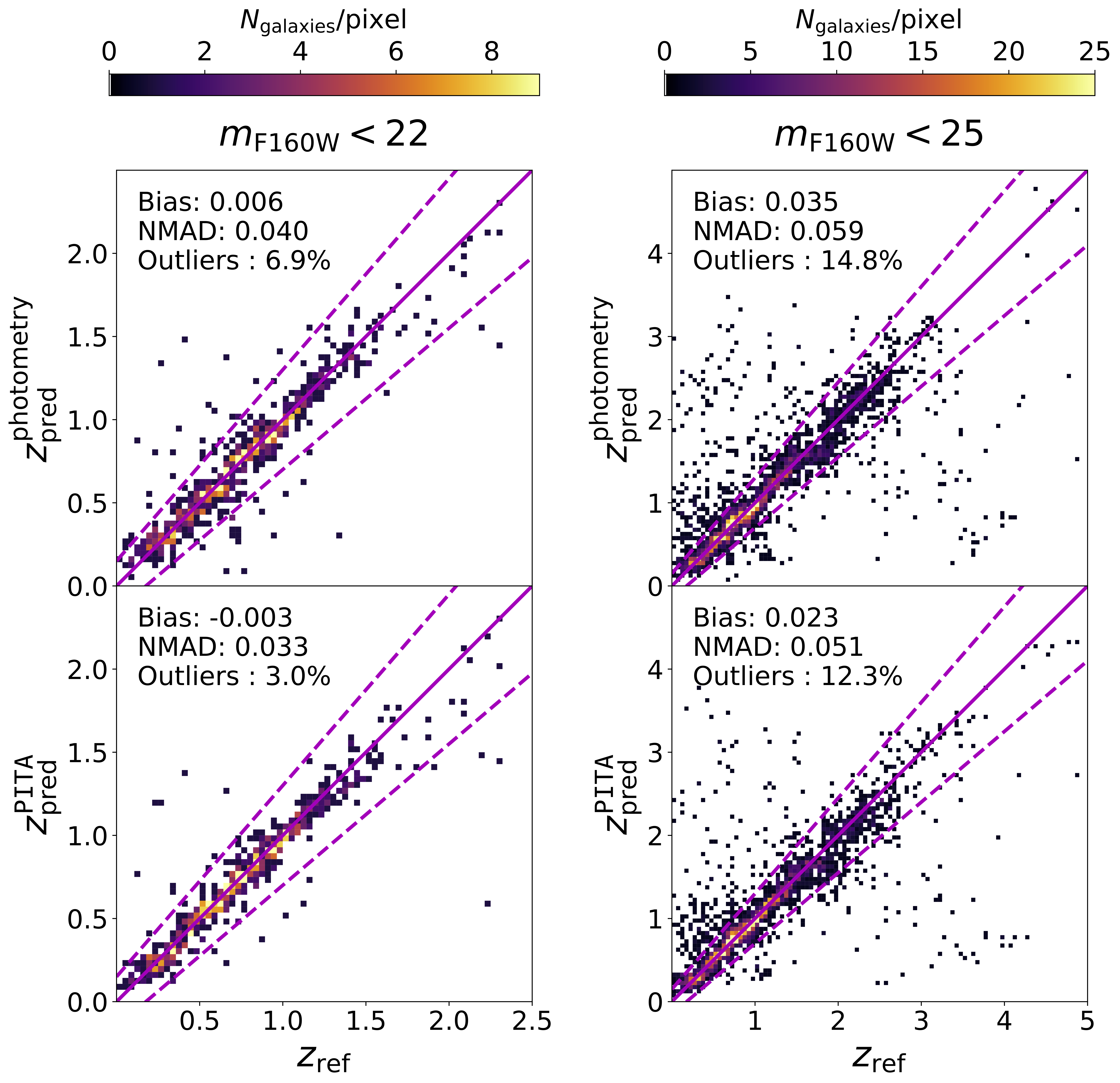}
    \caption{Predicted vs.\ reference redshifts for galaxies in the test set. The top row shows predictions from photometry-MLP on the y-axis, while the bottom row shows predictions from \texttt{PITA}. The left column shows the 800 galaxies with $m_\mathrm{F160W} < 22$ (out to $z\sim2.5$), and the right column shows the 2,911 galaxies with $m_\mathrm{F160W} < 25$ (out to $z\sim4$). Compared to the photometric baseline, the semi-supervised method reduces bias by $50\%$, $\sigma_{\text{NMAD}}$ by $18\%$, and $f_{\text{outlier}}$ by $57\%$ for the bright subsample. The gains in $\sigma_{\text{NMAD}}$ and $f_{\text{outlier}}$ are visually evident in the tighter clustering of points around the one-to-one line.}
    \label{fig:pred_vs_true_combined}
\end{figure*}

Our deep, fully space-based CANDELS dataset provides an opportunity to answer the central question: can deep learning applied directly to resolved galaxy images outperform photometry-only photo-$z$'s in a Roman-like dataset? To investigate this, we calculated several standard performance metrics widely-used in the photo-$z$ literature to compare the results from our deep learning algorithms to those from photometry-only predictions.

We further examined the mapping of latent vectors to quantities of interest in an attempt to investigate the underperformance of the self-supervised method and the success of \texttt{PITA}. Given that contrastive methods have yielded promising results on low-redshift datasets, such as DESI or Sloan Digital Sky Survey (SDSS; \citealt{2000AJ....120.1579Y}), this can shed light on why they may be less effective at higher redshifts and how to optimize photo‑$z$ estimation strategies for Roman's imaging-rich but label-scarce datasets.

\subsection{Photo-$z$ Performance}
\subsubsection{Photo-$z$ Performance Metrics} \label{performance_metrics}
To evaluate the performance of each method, we calculated a number of statistical metrics that are standard in the photo-$z$ literature. Based upon the normalized difference between the reference and predicted redshift, $\Delta{z} = \frac{z_{\text{ref}}-z_{\text{pred}}}{1+z_{\text{ref}}}$, we computed the following:
\begin{itemize}
    \item the \textbf{bias} $\langle \Delta{z} \rangle$, defined as the mean of $\Delta{z}$, measures the average value of the prediction error;
    \item the \textbf{normalized median absolute deviation (NMAD)}, defined as
        \begin{equation*}
            \sigma_{\text{NMAD}} = 1.4826\times\text{Median}\Big(|\Delta{z} - \text{Median}(\Delta{z})|\Big),
        \end{equation*}
        which provides a robust estimate of the spread of prediction errors; and
    \item the \textbf{fraction of Outliers $f_{\text{outlier}}$}, defined as the fraction of objects that have $|\Delta{z}|>0.15$, which assesses the rate of catastrophic failures.
\end{itemize}

\noindent We calculated these metrics using a test set of 3,094 objects that was separate from the training and validation sets.

% \begin{table*}
%     \centering
%     \begin{tabular}{lcccccc}
%     %\hline
%     {} & \multicolumn{3}{c}{Test Set} & \multicolumn{3}{c}{$m_{\text{F160W}}<22$} \\
%     %\hline
%     Method & Bias & $\sigma_{\text{NMAD}}$ & $f_{\text{outlier}}$ & Bias & $\sigma_{\text{NMAD}}$ & $f_{\text{outlier}}$ \\
%     \hline
%     Photometry & 0.045 & 0.064 & 17.7\% & 0.012 & 0.040 & 5.6\% \\
%     %\hline
%     Supervised & 0.033 & 0.061 & 15.6\% & 0.007 & 0.039 & 3.9\% \\
%     %\hline
%     Fine-tuned Self-supervised & 0.043 & 0.070 & 17.7\% & 0.014 & 0.046 & 5.9\% \\
%     %\hline
%     Semi-supervised & \textbf{0.029} & \textbf{0.054} & \textbf{15.3\%} & \textbf{0.002} & \textbf{0.032} & \textbf{3.0\%} \\
%     %\hline
%     \end{tabular}
%     \caption{Performance metrics (Bias, $\sigma_{\text{NMAD}}$, and $f_{\text{outlier}}$) for each method calculated using all sources in the test set (left column) and sources brighter than {$m_{\text{F160W}}=22$} (right column). The values for the best performing method are highlighted in bold.}
%     \label{tab:metrics_table}
% \end{table*}

\subsubsection{Comparisons of Photo-$z$ Performance}
Table \ref{tab:metrics_table} shows the values of these performance metrics (bias, $\sigma_\mathrm{NMAD}$, and $f_\mathrm{outlier}$) for the different methods calculated using the 800 galaxies in the test set that had $m_\mathrm{F160W}<22$  (left) or the 2,911 galaxies with $m_\mathrm{F160W}<25$ (right). For bias, values closer to 0 are better, while for $\sigma_\mathrm{NMAD}$ and $f_\mathrm{outlier}$ smaller values are preferred. In each column the best values are highlighted in bold.

We find that in this analysis the LePHARE point estimates (modes of the redshift posteriors $P(z)$) exhibit significantly worse bias, scatter, and number of outliers compared to all the ML approaches. Template-fitting approaches are heavily dependent on well-calibrated galaxy SED templates, which would require deep, multi-wavelength spectroscopic observations at the depths and redshifts of our dataset to optimize; on the other hand, they are likely to extrapolate better than ML methods beyond the bounds of existing training sets. It should be noted that we used the default \texttt{LePHARE} hyperparameters in the RAIL package. Given that it has performed well in other cases \citep{2020A&A...644A..31E}, we expect that extensive hyperparameter tuning can potentially improve results.

For the deep learning algorithms, our first finding is that the fully-supervised algorithm outperforms both the template-based and MLP-based photometry-only methods, with significant improvements in almost all metrics. This demonstrates that the pixel-level data in CANDELS imaging contains additional redshift information compared to integrated photometry and so can improve photo-$z$ estimates for individual galaxies.  Given the similarities between CANDELS and anticipated Roman data, we would expect improvements over photometry-based methods for Roman as well. In addition, with the wider areas covered by Roman fields, we expect to have many more spectroscopic labels for bright Roman sources compared to what is available in CANDELS, which likely would lead to further improvements in performance. Photo-$z$'s for faint sources, however, will likely remain label-limited. In Section \ref{subsec:scaling}, we examine how the performance of each method scales with the size of the labeled training set and discuss implications for Roman.

Given the limited number of spectroscopic labels in our dataset and the similar constraints expected for Roman (particularly for faint sources), we trained a self-supervised contrastive learning algorithm in an attempt to leverage unlabeled data and produce even better estimates. Despite this approach working well at low redshift with ground-based imaging, its performance in this application was comparable to the photometry-MLP method and notably worse than the fully-supervised algorithm. We explore potential causes of this behavior in Section \ref{subsec:umap}. 

Finally, we found that our semi-supervised \texttt{PITA} algorithm achieved the best performance across almost all evaluated metrics. Compared to the photometry-MLP baseline, it reduced bias by $34\%$, $\sigma_{\text{NMAD}}$ by $14\%$, and $f_{\text{outlier}}$ by $17\%$. The improvements are even greater for bright galaxies ($m_{f160w} < 22$), with reductions of $50\%$ in bias, $18\%$ in $\sigma_{\text{NMAD}}$, and $57\%$ in $f_{\text{outlier}}$. This is presumably because brighter galaxies have high S/N images and more robust redshift labels, as the grism-$z$'s and COSMOS2020 many-band photo-$z$'s which dominate at fainter magnitudes are less reliable. Since Roman will have $\sim$ three orders of magnitude more galaxy images than CANDELS and correspondingly more spectroscopic labels for bright sources, we would expect greater improvements over the photometry-only baseline in Roman analyses.

Figure \ref{fig:pred_vs_true_combined} shows the photo-$z$ predictions vs.\ the reference redshift labels for the photometry-MLP (top) and \texttt{PITA} (bottom). The points are binned into 2D histograms, and the color coding shows the number of points within a bin (pixel). The improvements are visually apparent in these scatter plots: the semi-supervised \texttt{PITA} predictions scatter less about the 1-to-1 line (indicating lower $\sigma_\text{NMAD}$), have a lower bias, and exhibit fewer catastrophic outliers. Notably, the MLP fails to predict any redshifts greater than $z\sim3$, whereas \texttt{PITA} successfully predicts those values roughly $50\%$ of the time despite the very small numbers of objects at those redshifts in the training set (cf. Figure \ref{fig:data_distributions}).

In the right column of Figure \ref{fig:pred_vs_true_combined}, a noticeable ``stair" pattern appears in the redshift range $z_{\text{ref}}=[1.5,2]$. We hypothesize that this is caused by the Balmer/$4000$ $\AA$ breaks, which are important features for photo-$z$ prediction but fall within the band gap between the F814W and F125W filters at these redshifts (see Figure \ref{fig:transmission_curves}). As a result, F814W $-$ F125W color remains approximately constant in this regime, increasing degeneracies and reducing redshift sensitivity. Photo-$z$ predictions using Roman's filters are not expected to suffer from this issue, as its filter bandpasses are more contiguous than CANDELS'.

\subsection{Visualization of Latent Vectors} \label{subsec:umap}
While deep learning algorithms are typically considered as ``black boxes'' with limited interpretability, it is nonetheless possible to probe the information content in the latent spaces produced by our algorithms. To that end, we used the Uniform Manifold Approximation and Projection (UMAP; \citealt{2018arXivUMAP}) algorithm to visualize the correlation of self-supervised and \texttt{PITA} latent vectors with redshift, magnitude, and color. UMAP is a non-linear dimensionality reduction algorithm that preserves local and global topological structure. We applied it to project the 128-dimensional latent vectors into 2D UMAP embeddings. This allowed us to investigate how effectively the vectors encoded relevant information using scatter plots color-coded by the properties of interest.

For each of the deep learning strategies tested, we trained a separate UMAP on the $\sim20,000$ labeled data using the Python package \texttt{UMAP} \citep{mcinnes2018umap-software}. The key hyperparameters for this code are the distance metric, the number of neighbors used, and the minimum distance between points in the low-dimensional space. We used the cosine distance metric with 50 nearest neighbors and a minimum distance of $0.2$.

\begin{figure*}
    \centering
    \includegraphics[width=1\linewidth]{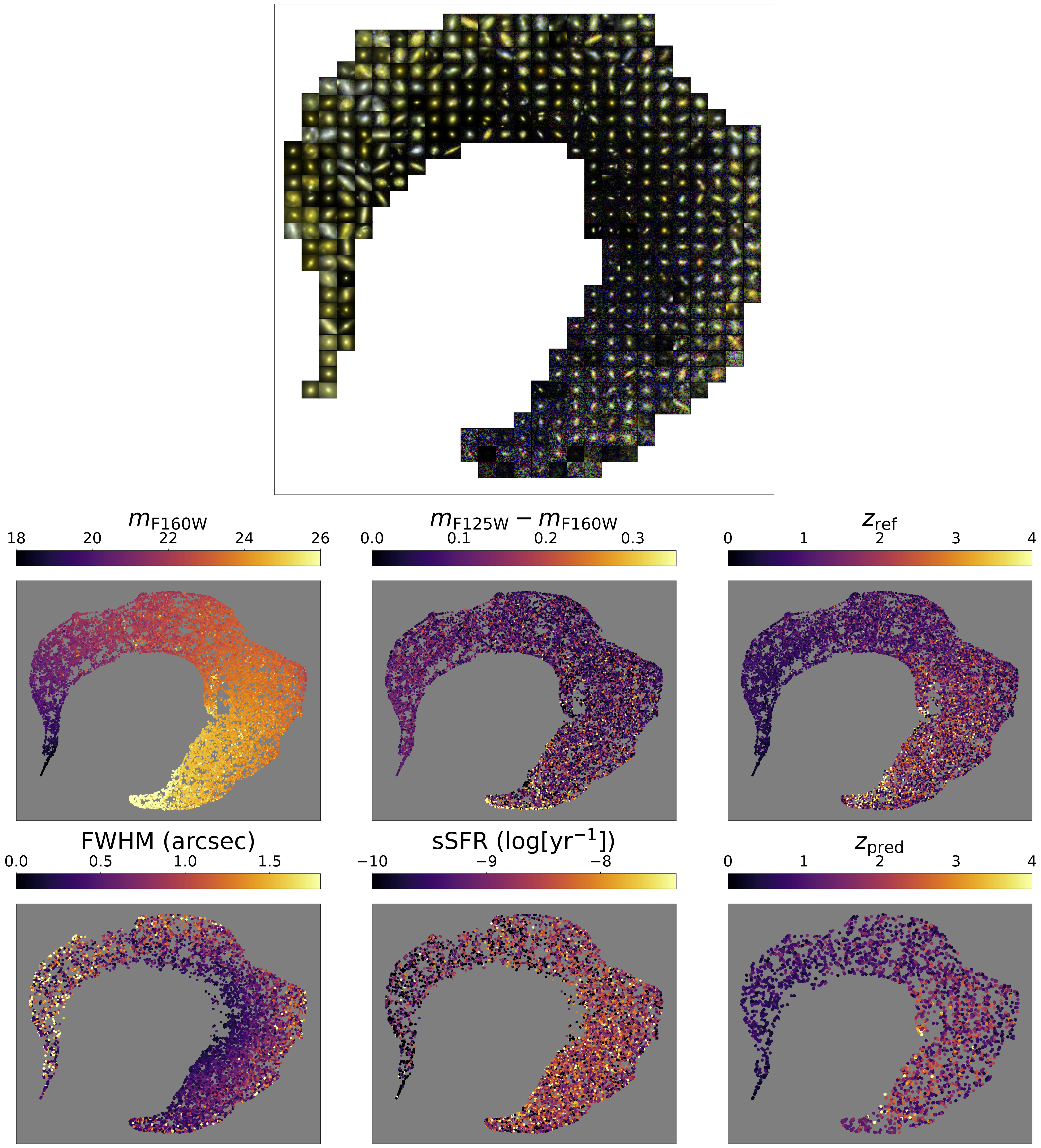}
    \caption{Two-dimensional UMAP embeddings of the self-supervised latent space. In the top panel, we divide the UMAP-derived space into a $30\times30$ grid, select a random galaxy from each grid cell, and display its image cutout. Galaxies in adjacent grid cells exhibit similar shapes and brightnesses but not necessarily similar colors. The middle panels show the variations of $m_\mathrm{F160W}$, F125W $-$ F160W color, and $z_\mathrm{ref}$ across the embedding space; the bottom panels show the variations of F814W FWHM, LePhare sSFR, and redshift predictions for the test set after fine-tuning. FWHM and sSFR were only available for COSMOS sources. While there are clear perpendicular trends of $m_\mathrm{F160W}$ and FWHM across the UMAP embedding space, this is not the case for redshift, color, or sSFR. These trends should reflect what information is effectively encoded in the self-supervised latent space. Interpolation in this space produces photo-$z$'s that are significantly biased and have large variance, even after fine-tuning, due to the lack of correspondence between latent space values and redshift.}
    \label{fig:self_supervised_umap_plots}
\end{figure*}

\begin{figure*}
    \centering
    \includegraphics[width=1\linewidth]{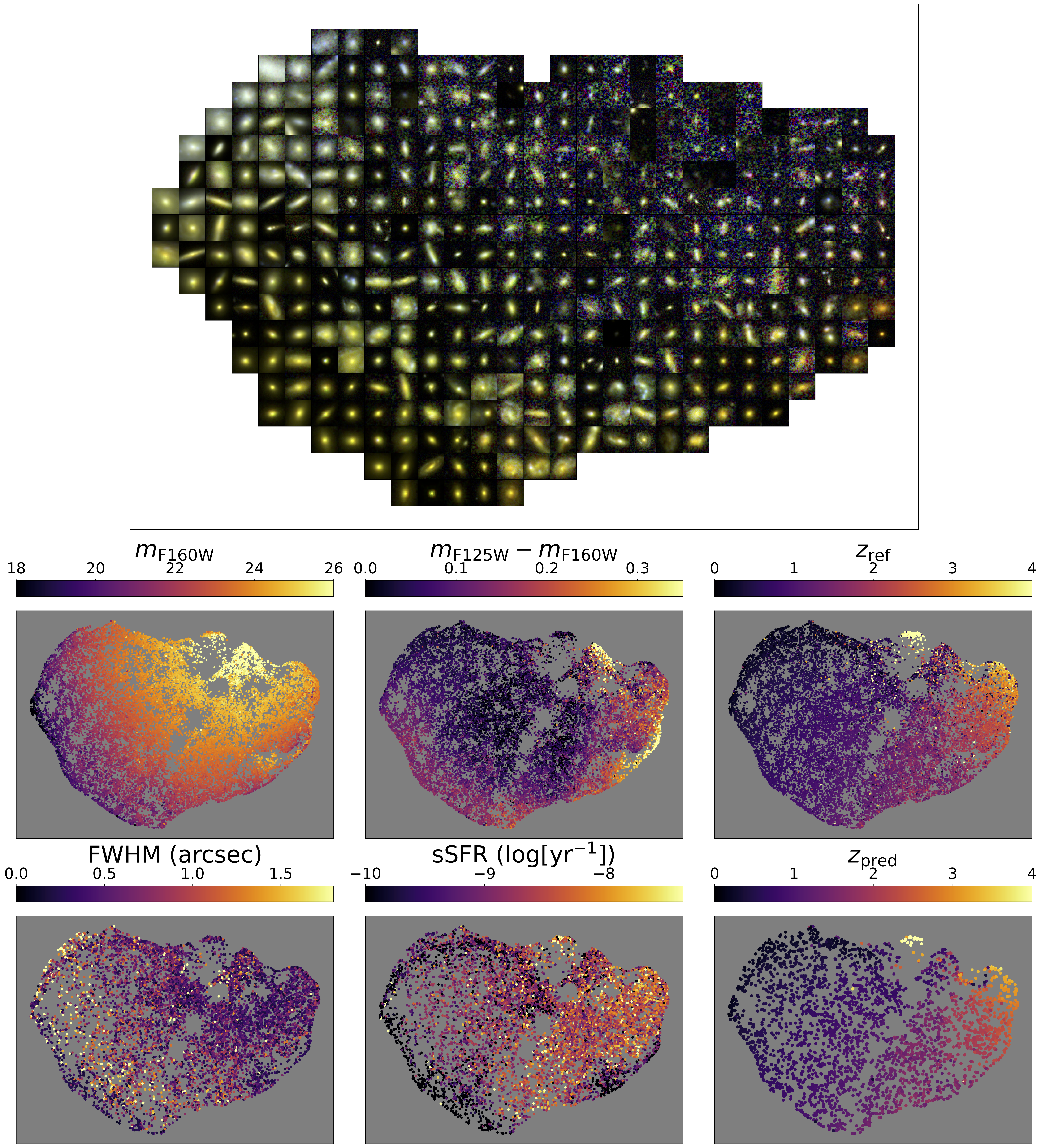}
    \caption{Similar to Figure \ref{fig:self_supervised_umap_plots}, but in this case applying UMAP to the representations obtained from \texttt{PITA}. In addition to encoding brightness ($m_\mathrm{F160W}$) and size (FWHM), these representations also show much smoother trends of redshift, color, and sSFR across the latent space, enabling improvements in photo-$z$ estimates from latent vectors.}
    \label{fig:semi_supervised_umap_plots}
\end{figure*}

\subsubsection{Self-supervised Latent Vectors}
Figure \ref{fig:self_supervised_umap_plots} shows results based on the 2D UMAP embeddings of the self-supervised algorithm's latent vectors. In the top panel, we divide the UMAP space into a grid and visualize the types of galaxies occupying each region by showing image cutouts of randomly selected examples. The middle panels show trends with $m_\mathrm{F160W}$, F125W $-$ F160W color, and reference redshift ($z_\mathrm{ref}$). The bottom left and middle panels show trends with FWHM (measured assuming a Gaussian core in F814W) and specific star formation rate (sSFR; from multi-band LePHARE template fitting) for objects in the COSMOS field from the COSMOS2020 catalog \citep{2022ApJS..258...11W}. The bottom right panel shows the predicted redshifts for the test set after fine-tuning ($z_\mathrm{pred}$).

The top panel of Figure \ref{fig:self_supervised_umap_plots} shows that the embeddings form a sequence of increasing magnitude that runs clockwise around the arc (which can also be seen in the $m_\mathrm{F160W}$ panel) and a roughly orthogonal sequence of size increasing radially outward (also seen in the FWHM panel). However, the trend in photometric color is not well ordered across this space. Blue galaxies neighbor yellow and red ones in the top panel;  this lack of color separation can also be seen in the $m_\mathrm{F125W} - m_\mathrm{F160W}$ panel. As a result, sSFR and redshift, which are strongly correlated with color, are also not well ordered. This is true especially at higher redshifts, as can be seen in the $z_\mathrm{ref}$ and sSFR panels. While this trend is improved after fine-tuning, we still see in the $z_\mathrm{pred}$ panel that the predicted redshifts do not exhibit a smooth gradient, but instead there is large scatter in redshift within small neighborhoods in embedding space.

The observed trends in the UMAP embeddings reflect the information encoded in the self-supervised latent vectors. These vectors appear to primarily encode magnitude and size, rather than color, redshift, or sSFR, leading to suboptimal photo-$z$ predictions. Based on this finding, we developed the \texttt{PITA} algorithm to combine the contrastive loss with explicit objectives for predicting color and redshift, thereby guiding the latent space toward representations that are more informative for the downstream photo-$z$ task.

\subsubsection{\texttt{PITA} Latent Vectors}
Figure \ref{fig:semi_supervised_umap_plots} shows the 2D UMAP embeddings of the \texttt{PITA} latent vectors. In the top panel, we see that galaxies near each other in embedding space have similar morphologies, colors, and brightnesses. This is reflected in the smaller panels, which exhibit clear and smooth trends in magnitude, color, FWHM, sSFR, and redshift.
However, the dependence on color is not monotonic. This appears to be an artifact of projecting onto a 2D UMAP; we have found that 3D UMAP embeddings display a more monotonic behavior, but for visual clarity we show only the 2D embeddings in this paper.

The smooth gradients of color, sSFR, and redshift across embedding space indicate that the color and redshift losses are properly aligning the \texttt{PITA} latent vectors to encode galaxy SED information. The fact that there is still a clear trend with size (FWHM) indicates that, while the redshift and color losses help to align the latent space, the contrastive loss still encourages the latent vectors to encode morphological information as well as SED properties. 

The $z_\mathrm{ref}$ panel reveals the presence of outliers --- i.e., low-redshift galaxies surrounded by high-redshift neighbors, or vice versa. These may correspond to incorrect redshift measurements (mostly attributable to grism and COSMOS2020 photo-$z$'s) or instead to cases where the 4-band imaging lacks sufficient information to resolve redshift degeneracies. For these cases, the output of \texttt{PITA}, as shown in the $z_\mathrm{pred}$ panel, tends to reflect the average redshift of neighboring galaxies. This is a positive sign that the model is not overfitting. Overall, the predictions successfully capture the gradient of redshift across the embedding space.

\begin{figure*}
    \centering
    \includegraphics[width=1\linewidth]{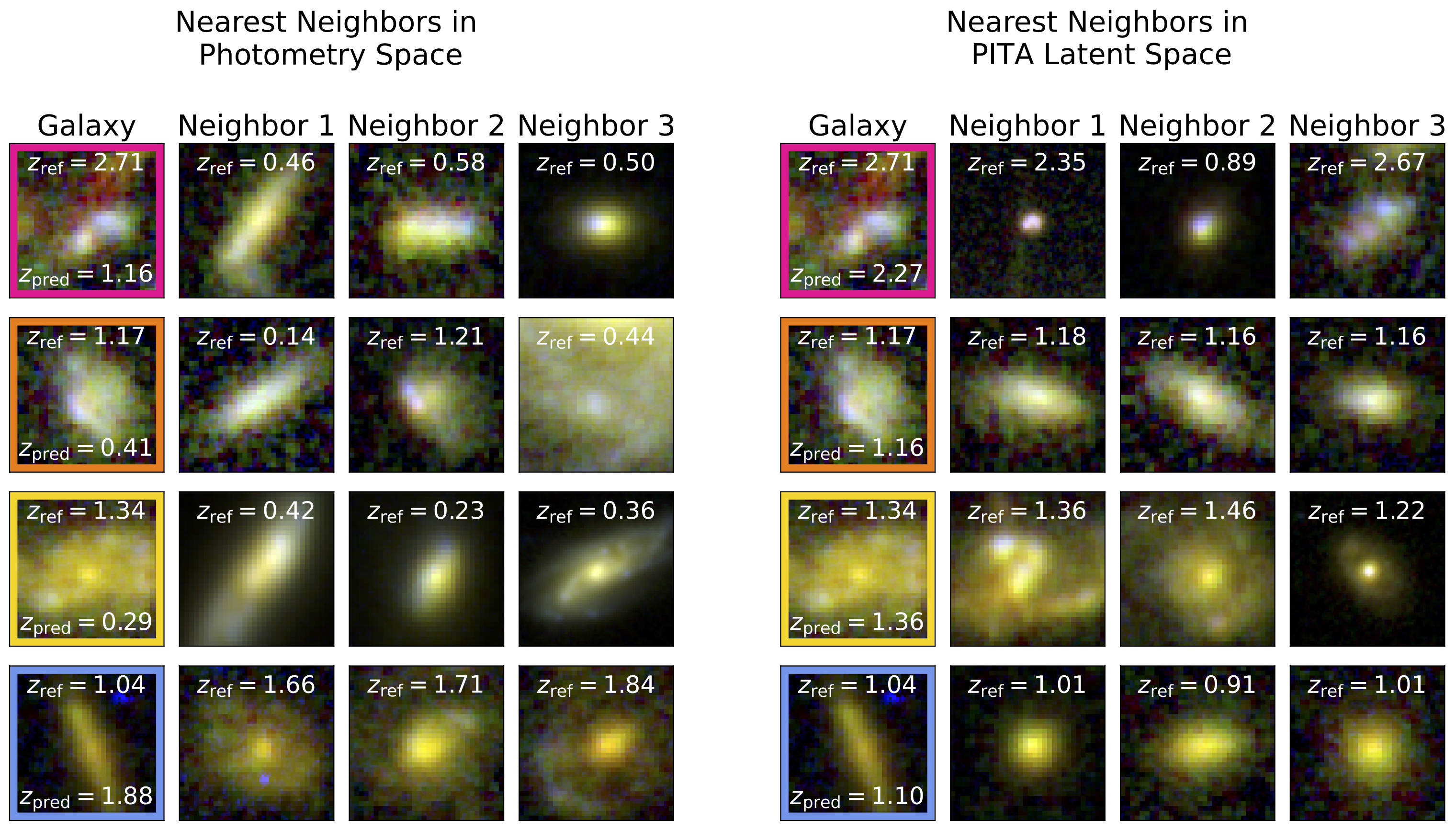}
    \caption{RGB images of four galaxies from the test set and their nearest neighbors in photometry space (left) or in the \texttt{PITA} latent space (right). The four images with colored square boxes are galaxies that have catastrophically wrong photo-$z$ predictions from photometry ($|\Delta{z}|>0.15$) but correct predictions from \texttt{PITA} ($|\Delta{z}|<0.15$). The first three rows suggest that \texttt{PITA} leverages morphological information to break color-redshift degeneracies and improve photo-$z$ estimates, even for galaxies affected by imaging artifacts (first row). However, the last row shows an example where \texttt{PITA} makes a correct prediction even when the neighbors do not look morphologically similar, suggesting that \texttt{PITA} is leveraging pixel-level information beyond just integrated photometry and morphology.}
    \label{fig:examples_for_neighbors}
\end{figure*}

\subsection{Breaking Color--redshift Degeneracies} \label{subsec:breaking_degeneracies}
Figure \ref{fig:examples_for_neighbors} shows four examples of galaxies from the test set that have catastrophically wrong predictions from photometry but correct predictions (within $15\%$) from \texttt{PITA}. For the first three galaxies, the neighbors in photometry space have similar colors but not necessarily similar morphologies or redshifts. The neighbors in the \texttt{PITA} latent space, in contrast, have comparable redshifts and morphologies as well as similar colors, resulting in better predictions. These results support the most common hypothesis for explaining the superior performance of deep learning photo-$z$ algorithms at low redshift: predictions can improve because morphology can help to break degeneracies in the color--redshift relation. However, the last row shows an example where \texttt{PITA} makes a correct prediction even when the neighbors do not look morphologically similar, suggesting that the algorithm is leveraging pixel-level information in more complicated ways. We discuss the implications of these findings in Section \ref{subsec:discussion_photometry_vs_deep_learning}.

Interestingly, the photometric nearest neighbors of the first galaxy appear more yellow in their image cutouts, whereas the primary galaxy itself has subregions with colors ranging from blue to pink. This discrepancy may be due to the photometry pipeline failing to completely separate the diagonal red streaks coming from an outside source from the flux of the object of interest, resulting in redder photometric colors overall. In contrast, the \texttt{PITA} nearest neighbors exhibit internally varying blue-to-pink colors that more closely match the primary galaxy, suggesting that the learned latent vectors are more robust to contamination and better capture intrinsic galaxy features. This behavior may also improve deblending, particularly in joint analyses that combine ground- and space-based imaging (e.g., from Roman and LSST). While blending is an important issue for future surveys \citep{2021NatRP...3..712M}, a detailed investigation of its impact is beyond the scope of this work.

\subsection{Scaling of Performance with the Size of Redshift Training Sets} \label{subsec:scaling}
\begin{figure*}
    \centering
    \includegraphics[width=1\linewidth]{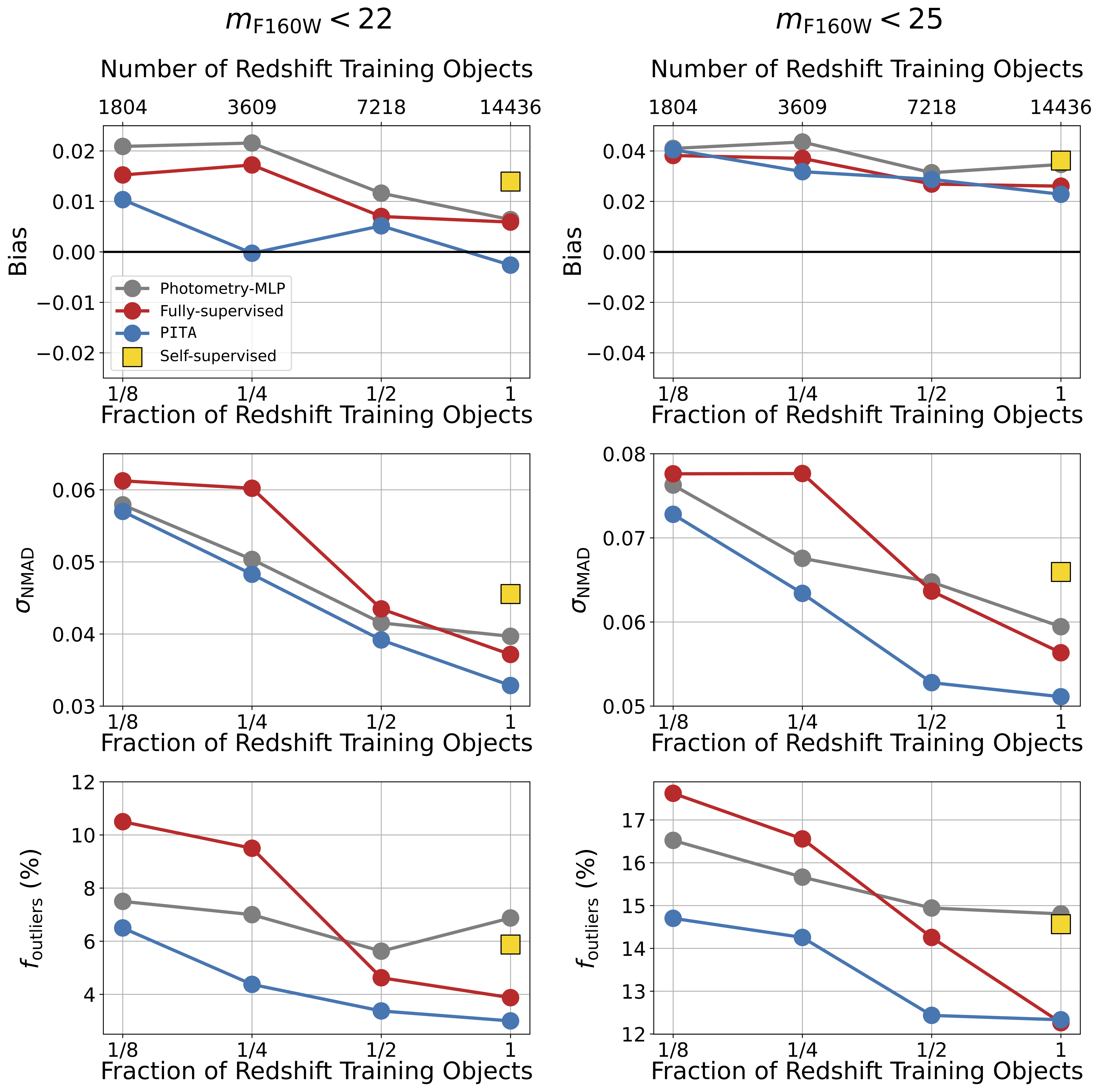}
    \caption{Scaling of performance metrics (bias, $\sigma_{\text{NMAD}}$, and $f_{\text{outlier}}$) as a function of the number of labeled objects used in training the photometry-MLP, fully-supervised, and \texttt{PITA} models, and in fine-tuning the self-supervised model. The metrics are evaluated on test set sources with $m_\text{F160W} < 22$ (left) or with $m_\text{F160W} < 25$ (right). For next-generation imaging surveys (including Roman), the number of redshift labels will decrease drastically at greater magnitudes; crucially, \texttt{PITA} outperforms all methods regardless of the number of labels, but especially so when training sets are sparse, as will be true in the most difficult domains.} 
    \label{fig:spec_scaling_inc_supervised}
\end{figure*}

Figure \ref{fig:spec_scaling_inc_supervised} compares the performance of our semi-supervised \texttt{PITA} algorithm to the remaining ML methods as a function of the size of the redshift training set used. The training data points were removed randomly and the same remaining sources were used to train all the methods. Across the board, performance improved as the size of the labeled training set increased, consistent with expectations for ML models. This highlights the potential impact of obtaining greater numbers of spec-$z$ labels for future surveys (\citealt{2015APh....63...81N}; \citealt{2022arXiv220401992B}; Dey et al.\ in preparation). The Figure also shows the values of the self-supervised algorithm after fine-tuning with all the available objects in the training set (yellow squares). Since the performance was subpar, we did not attempt to fine-tune with fewer objects.

Interestingly, some trends in the photometry-MLP metrics appear to weaken beyond $\sim10,000$ labeled data points. While additional training data typically improves ML predictions, this behavior suggests that the potential for further gains in the photometry-MLP case is limited. In contrast, despite exhibiting poor performance when there are very few labels, the fully-supervised algorithm scaled better and outperformed the photometry-MLP algorithm for the largest training sets. This is not surprising, as deep learning CNNs are data-hungry but outperform classical ML in many applications when provided with enough training data. 

However, \texttt{PITA} outperformed all other methods across all performance metrics, even when very few redshift labels were used. This is because the contrastive and color losses take advantage of unlabeled data to learn informative latent vectors, and the small subset of labeled examples is then used to better align those vectors with a redshift loss.  The \texttt{PITA} latent space is therefore well-suited to interpolation between sparse training redshifts, allowing optimal use of samples that are limited in size.

\section{Discussion} \label{sec:discussion}
Deep learning algorithms that can learn from unlabeled data are advancing at a rapid pace and are becoming increasingly popular. Previous work has already shown the effectiveness of these models in photo-$z$ estimation, and they offer a promising avenue for obtaining the best photo-$z$'s for Roman by leveraging vast amounts of unlabeled images.

Our HST CANDELS dataset enabled us to test such algorithms in a Roman-like regime that is radically different from previous datasets, with spatially-resolved galaxies up to $m_\mathrm{F160W}\sim25$ and $z\sim3$. In this section, we discuss the results of comparing fully-, self-, and semi-supervised algorithms to photometry-only methods. We speculate on the reason behind deep learning's success, discuss the possibilities of leveraging unlabeled data (including using foundation models), and the advantages of \texttt{PITA} for the specific task of photo-$z$ prediction from Roman's images. 

\subsection{Photometry-based Versus Deep Learning Photo-$z$'s} \label{subsec:discussion_photometry_vs_deep_learning}
In Section \ref{sec:results} we showed that deep learning algorithms applied directly on images outperformed photo-$z$'s predicted from photometry alone for a Roman-like dataset.
%, with our semi-supervised PITA algorithm performing the best by leveraging unlabeled data. 
However, the question of \textit{why} pixel-level information improves photo-$z$'s remains an open one. 

Figure \ref{fig:examples_for_neighbors} shows that while galaxy morphology may play a role, nearest neighbors in the \texttt{PITA} latent space are not always visually similar; the exact contribution of morphology remains unclear. Recently, \citealt{2026ApJ...997...36M} investigated the common regression phenomenon of attenuation bias \citep{2025OJAp....8E..95T} within the context of deep learning photo-$z$'s. This effect causes the regression predictions from noisy inputs to be biased towards the mean (with high values being under-predicted and low values over-predicted). Using SDSS data, they found that when compared to deep learning methods, photometry-based classical ML methods demonstrated substantially larger attenuation bias in local regions of color space. These local attenuation biases combine to give classical methods a higher scatter when assessed globally. They hypothesized that deep learning algorithms are learning the optimally weighted combinations of the many pixel-level SEDs of an object to predict its redshift. This suggests that certain pixels may be more informative about redshift than the relative weight that is attributed to them from the photometric model, thus allowing deep learning algorithms to outperform photometry-based ones.

Despite the challenges in interpreting the source of these improvements, the scalings shown in Figure \ref{fig:spec_scaling_inc_supervised} provide a strong case for using \texttt{PITA} or a similar algorithm to estimate photo-$z$'s from Roman imaging (and other future surveys). For bright galaxies with abundant redshift labels, the superior scaling of deep learning methods insures that they will provide better results than photometry alone. For faint galaxies for which labeled data are scarce, a semi-supervised approach that exploits the full set of objects can provide better results than either photometry-only or fully-supervised CNN methods.

\subsection{Impact of Exploiting Unlabeled Data}
The difference in scaling between the fully-supervised and \texttt{PITA} algorithms can be understood by viewing deep learning as a combination of feature extraction and regression. Initial layers learn to extract appropriate features from images, while the final layers perform a simple regression from those features to estimate redshift. In the fully-supervised case with less than $\sim10^4$ training data points, there is insufficient data for the algorithm to both accurately extract features and optimize regression results. In contrast, \texttt{PITA} uses the large set of unlabeled images to learn informative features, while exploiting the much smaller set of redshift labels to train the regression, leading to improved performance in both label-scarce and label-rich regimes. Although simplistic, this model offers an intuitive explanation of the observed behaviors and highlights the importance of learning informative features from large unlabeled imaging datasets.

\subsubsection{The Advantages of \texttt{PITA}}
The vast majority of sources detected by Roman (or other next-gen photometric surveys) will not have spec-$z$ labels, especially at high redshifts and faint magnitudes. Therefore, self- and semi-supervised algorithms that can leverage unlabeled data to learn informative features from images will be extremely valuable.

One of the primary goals of this work was to determine if self-supervised contrastive learning could yield a latent space based of deep, space-based images that encodes redshift information out to $z\sim3$. The UMAP projections in Figure \ref{fig:self_supervised_umap_plots} provide a qualitative means of answering this question. At low redshifts ($z < 0.3$), the smooth redshift gradient indicates that the representations effectively capture redshift. Indeed, previous findings have found that contrastive pre-training improved photo-$z$ estimates in this regime. It is only beyond $z \sim 0.3$ that we begin to observe neighboring points in latent space with significantly different redshifts. This is likely driven by the stronger correlation of redshift with brightness at low redshifts where the luminosity distance changes rapidly \citep{1934rtc..book.....T}. Since contrastive learning effectively encodes brightness, the resulting representations are particularly informative for predicting low redshifts, especially after fine-tuning.

Ultimately, the contrastive representations encode the information necessary to minimize the InfoNCE loss function introduced in Eq.~\ref{eq:contrastive_loss}. This can be viewed through the lens of dimensionality reduction. To minimize the loss function and separate positive pairs from negative ones, the network compresses high-dimensional image data into a latent space that captures the variance in the images. For our deep, space-based imaging dataset, the dominant sources of variance in pixel values are brightness and morphology; these are therefore more easily captured by the model. As opposed to the orders-of-magnitude variation in brightness, the dynamic range of galaxy colors is comparatively smaller, making it a higher-order source of variance. As a result, the contrastive learning latent vectors primarily capture the first-order properties but are worse at including the effects of higher-order properties such as color and redshift; in other words, encoding brightness and morphology is enough to separate positive pairs from negative ones. In contrast, in SDSS-like datasets of bright, low-$z$ galaxies with ground-based imaging, we suspect that colors represent a significantly larger fraction of the variance in the data. This would explain why contrastive methods are better able to encode colors and redshifts in the latent space in that domain, however further investigation is required to test this hypothesis. It is also the case that at low redshifts, brightness is more strongly correlated with redshift, and morphology is more strongly correlated with observed color (due to it being more strongly correlated with rest-frame color), thus contrastive methods can be implicitly encoding color and redshift information in the low-redshift regime.

\texttt{PITA} was designed to avoid these limitations by jointly optimizing for contrastive, color prediction, and redshift prediction losses. This encourages the latent space to encode features that are both morphologically and photometrically informative for redshift prediction. As seen in Figure \ref{fig:semi_supervised_umap_plots}, the \texttt{PITA} latent space varies smoothly with color and redshift, enabling the best photo-$z$ estimates in both label-scarce and label-rich regimes.

\subsubsection{Comparison to Other Strategies}
The deluge of data from Roman, Rubin, Euclid, DESI, and other surveys will enable training large multi-modal foundation models that integrate different instruments and data modalities into a shared latent space \citep{2024arXiv240514930S, 2025arXiv250315312E, 2025arXiv251017960P}. The goal for doing so is to obtain generalizable representations of the data that could serve a wide range of downstream tasks, from anomaly detection to classification and regression (including photo-$z$ estimation). However, our findings suggest that photo-$z$ performance is sensitive to how the latent space is learned, particularly to the choice of the self-supervised loss function. In general, it might be challenging to identify a single self-supervised objective that yields optimal results across all downstream tasks, even after fine-tuning. In fact, previous studies have found that specialized uni-modal networks can outperform more general multi-modal ones in certain scenarios \citep{wang2020makes,2022arXiv220312221H}.

For the downstream task of photo-$z$ estimation, the contrastive learning model's tendency to prioritize brightness and morphology over color poses a significant limitation. Our results also highlight the importance of dataset- and downstream-task-specific pre-processing and augmentation strategies; certain features in the data may be more important than others for different downstream tasks, and in the case of contrastive learning, the choice of augmentations is critical in determining what information is retained \citep{tian2020makes}. A comprehensive exploration of contrastive strategies that are ideal for photo-$z$ inference is beyond the scope of this work, however.

Although our analysis focused specifically on contrastive learning, the insights gained may extend to other self-supervised tasks. When working with deep, space-based imaging datasets, it is important to realize potential shortcomings and use domain knowledge to obtain a more informative latent space. 

Generative models, which learn to reconstruct images from a latent space by minimizing a reconstruction loss, may encounter similar challenges. To first order, a low reconstruction loss can be achieved by reproducing the brightness across bands. However, small flux residuals (compared to the orders-of-magnitude dynamic range in flux) can result in large color residuals. This shortcoming could potentially be mitigated through pre-processing the data, weighting the loss function by brightness, or by explicitly including a color or redshift loss as we have done in this work, or as was done in \citealt{2022MNRAS.515.5285D}.

Transformer-based autoregressive models offer an alternative approach to resolving this issue. While an autoregressive reconstruction loss with images alone might not suffice, the inherent flexibility of transformers due to tokenization allows them to take as input both images and SEDs, thereby encouraging the latent space to encode color information, and redshift by extension \citep{2024arXiv240514930S,2025arXiv250315312E,2025arXiv251017960P}.

\subsection{Lessons for Roman}
Regardless of the specific self- or semi-supervised learning paradigm employed, the increased diversity of galaxy populations expected in upcoming surveys compared to current shallow, ground-based datasets calls for a cautious approach. Algorithms that perform well at low $z$ may struggle to generalize across the full range of galaxies that will be observed by Roman. 

Our primary goal in using unlabeled data is to learn informative low-dimensional latent vectors that encode images. However, the astronomical community has already developed a broadly used and highly efficient method for compressing images to a few numbers: integrated photometry. Photometric fluxes are strongly correlated with nearly every physical property of a galaxy, including its redshift \citep{2013ARA&A..51..393C}, and they can be measured for all images. This domain knowledge offers a powerful tool for assessing the performance of deep learning algorithms. Any latent space obtained by a deep learning algorithm should be at least as physically informative as photometry if it encodes all the photometric colors. However, pixel-level features in the images clearly include additional information, which can enable deep learning methods to outperform photometry-based ones.

\texttt{PITA} yields a physically informative latent space thanks to its triple loss function. By including color and redshift prediction losses, the algorithm is trained to learn latent vectors that can predict these features, thereby ensuring that it encodes them. This allows for smooth interpolation within the latent space, resulting in optimal photo-$z$ estimates for both the faint and bright sources that correspond to the label-scarce and label-rich regimes.

\section{Summary and Conclusions} \label{sec:summary_and_conclusions}
Scientific analyses with Roman's deep, high-resolution near-infrared imaging will heavily rely on photo-$z$'s. We tested deep learning photo-$z$ algorithms against traditional photometry-based methods on Roman-like data from the HST CANDELS survey. As a first step, we trained a fully-supervised CNN using $\sim20,000$ images with redshift labels and demonstrated that it outperformed traditional ML and template-fitting photometry methods. This confirmed that pixel-level information yields better photo-$z$'s even for faint galaxies at high redshifts.

Given Roman's high survey speed and wide field of view, spectroscopic follow-up will be highly incomplete, especially at high redshifts and faint magnitudes, leaving the vast majority of sources without spec-$z$ labels. We therefore also explored deep learning approaches that learn from unlabeled images. We pre-trained a CNN with $\sim80,000$ unlabeled CANDELS images via self-supervised contrastive learning, with the goal of learning low-dimensional latent vectors that encode redshift information. Then, we used objects with redshift labels to fine-tune the network for photo-$z$ prediction. However, contrary to previous results on low-redshift galaxies ($z<0.3$) with ground-based imaging, we found that self-supervised contrastive learning alone did not yield latent vectors that are optimal for redshift prediction at higher redshifts ($z>0.3$). 

To address this, we developed \texttt{PITA}: a semi-supervised algorithm that is simultaneously trained with images and photometric colors, as well as redshift measurements for the objects that have them. This is achieved by combining contrastive, redshift prediction, and color prediction loss functions. Unlabeled data (i.e., objects with or without $z$ measurements) contribute to the contrastive and color losses, while only labeled data contribute to the redshift loss, resulting in a latent space of images that is far more informative for downstream photo-$z$ inference. We showed that \texttt{PITA} outperforms all the other deep learning and photometry-based methods applied in all performance metrics used ($\sigma_\text{NMAD}$, $f_\text{outlier}$, and bias).

Our results highlight the importance of domain knowledge and multi-modal approaches. Recent work \citep{2024MNRAS.531.4990P} has shown that spectroscopic data can be used to obtain more informative latent vectors of multi-band imaging. However, deep, near-infrared-selected spectra will not be obtainable for the vast majority of Roman sources. We showed that in the absence of spectra, photometric colors computed from the images provide coarse SED information that can be used to achieve the same goal.

In Section \ref{subsec:scaling}, we further showed that \texttt{PITA} provides the best photo-$z$ estimates of the algorithms applied irrespective of the number of redshift training objects. This scalability is crucial: for both bright and faint sources, with redshift labels ranging from abundant to sparse, a semi-supervised approach is preferred in all cases.

As we push towards higher-redshift samples with deeper imaging, obtaining spectroscopic redshift labels will become increasingly difficult. Leveraging unlabeled objects will therefore become crucial for the next-generation of photometric surveys such as Roman. However, downstream photo-$z$ performance is sensitive to how the unlabeled objects are used. Methods like \texttt{PITA} that can effectively leverage the vast amounts of unlabeled data and that can exploit domain knowledge will yield the best possible photo-$z$ estimates, helping to advance studies focused on cosmology, galaxy evolution, and transient science.

\section*{Acknowledgments}
We would like to thank Anton Koekemoer, François Lanusse, Grant Merz, John F. Wu, Manuel Pérez Carrasco, Mike Smith, Marc Huertas-Company, and Rachel Bezanson for fruitful discussions.

A.~K., B.~H.~A., and J.~A.~N.\ acknowledge the support of the National Aeronautics and Space Administration under grant No.~80NSSC24K0082.

This research used resources of the National Energy Research Scientific Computing Center, a DOE Office of Science User Facility supported by the Office of Science of the U.S. Department of Energy under Contract No. DE-AC02- 05CH11231 using NERSC awards HEP-ERCAP0022859, HEP-ERCAP0033572, and HEP-ERCAP0036802.

A.\ K. thanks the LSST-DA Data Science Fellowship Program, which is funded by LSST-DA, the Brinson Foundation, the WoodNext Foundation, and the Research Corporation for Science Advancement Foundation; his participation in the program has benefited this work.

A.\ K. thanks the funding sources of the University of Pittsburgh Andrew Mellon and PITT PACC fellowships.

A.\ K. thanks AstroAI at the Center for Astrophysics $|$ Harvard \& Smithsonian, the California Institute of Technology, the Lawrence Berkeley National Laboratory, the University of Toronto, the University of California Los Angeles, and the Vector Institute for providing the support and opportunity to present this work and discuss with colleagues.

B. D. is a postdoctoral fellow at the University of Toronto in the Eric and Wendy Schmidt AI in Science Postdoctoral Fellowship Program, a program of Schmidt Sciences.

\textit{Software:} Numpy \citep{harris2020array}, Matplotlib \citep{Hunter:2007}, Scipy \citep{2020SciPy-NMeth}, Pandas \citep{reback2020pandas}, Astropy \citep{astropy:2013, astropy:2018, astropy:2022}, UMAP \citep{mcinnes2018umap-software}, Scikit-learn \citep{scikit-learn}, PyTorch \citep{2019arXiv191201703P}, PyTorch Lightning \citep{Falcon_PyTorch_Lightning_2019}, h5py \citep{h5py2008}, RAIL \citep{rail2025}.

\section*{Data Availability}
The code for PITA can be found on GitHub at \url{https://github.com/ashodkh/PITA} \citep{ashod_khederlarian_2026_19056551}.

The HST CANDELS imaging and photometric datasets used in this work are from MAST \citep{https://doi.org/10.17909/t94s3x}. Redshift labels were obtained from \citealt{2023ApJ...942...36K}, \citealt{2025arXiv250300120K}, and \citealt{2022ApJS..258...11W}. Additional properties for objects in the COSMOS field, such as FWHM and sSFR, were obtained from \citealt{2022ApJS..258...11W} (\dataset[10.26131/IRSA563]{https://www.ipac.caltech.edu/dois}).

The training, testing, and validation datasets we compiled from the above datasets are available on HuggingFace (\url{https://huggingface.co}). The \texttt{PITA} training dataset that contains both labeled and unlabeled images can be found at \dataset[doi:10.57967/hf/8044]{https://doi.org/10.57967/hf/8044}. The labeled training, validation, and test datasets (images and metadata) can be found at \dataset[doi:10.57967/hf/8045]{https://doi.org/10.57967/hf/8045}. These datasets were used for the fully-supervised algorithms, and the labeled validation and test datasets were also used to validate and test \texttt{PITA}.

\appendix

\section{Implementation of Contrastive Learning as a Dictionary Look-up Task} \label{app:contrastive_learning}
Our implementation of contrastive learning follows \citealt{he2020momentum}, where the algorithm is framed as a dictionary look-up task. In this framework, a set of $K$ keys $\{k_0,k_1,k_2,\dots\,k_{K}\}$ are stored in a dictionary, where each key $k_i$ corresponds to the latent representation of an augmented image $T(x_i)$. A representation $q$ (referred to as the query) of a given transformed image $T(x_j)$ has exactly one positive key $k_+$ in the dictionary that is derived from the same original image $x_j$, albeit through a different random application of $T$.\footnote{Since the transformations  are stochastic, each application of $T$ to an image $x_j$ produces a unique augmented version of that image.}

The loss function used for the query $q$ is the InfoNCE function \citep{oord2018representation}:
\begin{equation}
    l_{q} = -\text{log}\frac{\exp(S_C(\boldsymbol{q},\boldsymbol{k}_+)/\tau)}
    {\sum_{i=0}^{K}\exp(S_C(\boldsymbol{q},\boldsymbol{k}_i)/\tau)}.
\end{equation}
\noindent where $\tau$ is a temperature hyperparameter and the sum is over all the keys in the dictionary, including the positive key $k_+$. This loss function can be interpreted as the categorical cross-entropy loss function for classifying $q$ as being the positive pair of $k_+$. The total loss for a training batch is computed as the average of the losses $l_q$ for each query image. 

The query and key representations, $q=f_q(T(x))$ and $k=f_k(T(x))$, are obtained by applying CNN encoders $f_q$ and $f_k$ to the augmented images. In our implementation, we used ConvNeXt networks for this. A straightforward approach is to use a shared encoder for both queries and keys --- i.e., to have $f_k=f_q$. However, this can reduce the consistency of the key representations if the weights of the query network change rapidly during training. For this reason, we adopted the momentum update mechanism introduced in \citealt{he2020momentum}:
\begin{equation}
    \theta_{k} \leftarrow m\theta_k + (1-m)\theta_q
\end{equation}
where $\theta_k$ and $\theta_q$ are the parameters of the query and key encoders, respectively. We set the momentum coefficient to $m=0.999$, which leads to a slowly-evolving key encoder. The query encoder parameters $\theta_q$ are updated directly through back-propagation. We used a dictionary size of $K=60,000$ galaxies. Larger dictionaries did not lead to significant improvements.

\section{Impact of \texttt{PITA} Loss Weights} \label{app:loss_weights}
To combine the contrastive, color, and redshift losses, we first normalized the contrastive loss to have the same order of magnitude as the color and redshift losses ($10^{-2}$). Then, we tested different weighted sums:
\begin{equation}
    L_\mathrm{\texttt{PITA}} = w_{\mathrm{con}}L_{\text{contrastive}} + w_{\mathrm{color}}L_{\text{color}} + w_zL_{\text{z}},
\end{equation}
such that $w_{\mathrm{con}} + w_{\mathrm{color}} + w_z = 1$.

% \begin{deluxetable}{lccc}
%     \tablecaption{Photo-$z$ Performance With Different Weights \label{tab:weights_table}}
%     \tablehead{Weights & bias & $\sigma_\mathrm{NMAD}$ & $f_\mathrm{outlier}$}
%     \startdata
%     $w_z=w_\mathrm{con}=w_\mathrm{col}$ & \textbf{0.023} &  \textbf{0.051} & \textbf{12.3\%} \\
%     $w_z=10w_\mathrm{con}$, $w_\mathrm{col}=w_\mathrm{con}$ & 0.027 &  0.055 & 12.8\% \\
%     $w_z=100w_\mathrm{con}$, $w_\mathrm{col}=w_\mathrm{con}$ & \textbf{0.023} &  0.062 & 14.3\% \\
%     $w_\mathrm{col}=10w_\mathrm{con}$, $w_z=w_\mathrm{con}$ & 0.026 &  0.055 & 13.0\% \\
%     $w_\mathrm{col}=100w_\mathrm{con}$, $w_z=w_\mathrm{con}$ & 0.030 &  0.081 & 17.3\% \\
%     $w_z=w_\mathrm{con}/10$, $w_\mathrm{col}=w_\mathrm{con}$ & 0.027 &  0.058 & 13.5\% \\
%     $w_z=w_\mathrm{con}/100$, $w_\mathrm{col}=w_\mathrm{con}$ & 0.030 &  0.075 & 16.9\% \\
%     $w_\mathrm{col}=w_\mathrm{con}/10$, $w_z=w_\mathrm{con}$ & 0.029 &  0.057 & 13.3\% \\
%     $w_\mathrm{col}=w_\mathrm{con}/100$, $w_z=w_\mathrm{con}$ & 0.027 &  0.057 & 13.2\% \\
%     \enddata
%     \tablecomments{Photo-$z$ performance metrics (bias, $\sigma_{\text{NMAD}}$, and $f_{\text{outlier}}$) for different combinations of loss function weights. The best values are highlighted in bold and they correspond to equal weights.}
% \end{deluxetable}

\begin{deluxetable}{cccccc}
    \tablecaption{Photo-$z$ Performance With Different Relative Weights \label{tab:weights_table}}
    \tablehead{$w_\mathrm{con}$ & $w_\mathrm{color}$ & $w_z$ & bias & $\sigma_\mathrm{NMAD}$ & $f_\mathrm{outlier}$}
    \startdata
    1 & 1 & 1 & \textbf{0.023} &  \textbf{0.051} & \textbf{12.3\%} \\
    1 & 1 & 10 & 0.027 &  0.055 & 12.8\% \\
    1 & 1 & 100 & \textbf{0.023} &  0.062 & 14.3\% \\
    1 & 10 & 1 & 0.026 &  0.055 & 13.0\% \\
    1 & 100 & 1 & 0.030 &  0.081 & 17.3\% \\
    1 & 1 & $1/10$ & 0.027 &  0.058 & 13.5\% \\
    1 & 1 & $1/100$ & 0.030 &  0.075 & 16.9\% \\
    1 & $1/10$ & 1 & 0.029 &  0.057 & 13.3\% \\
    1 & $1/100$ & 1 & 0.027 &  0.057 & 13.2\% \\
    \enddata
    \tablecomments{Photo-$z$ performance metrics (bias, $\sigma_{\text{NMAD}}$, and $f_{\text{outlier}}$) for different combinations of relative loss function weights. The best values are highlighted in bold and they correspond to equal weights.}
\end{deluxetable}

Table \ref{tab:weights_table} shows the photo-$z$ performance metrics on test set objects with $m_\mathrm{F160W} < 25$ resulting from different sets of weight values. We found that equal weights resulted in the best metric values. Increasing $w_\mathrm{col}$ degraded performance significantly more than decreasing it. In contrast, decreasing $w_z$ degraded performance more than increasing it did. This suggests that while \texttt{PITA} benefits from both the redshift and color losses, the redshift loss has greater impact on the final results. This is not surprising, as the downstream task we are using for this test is photo-$z$ estimation. Notably, it might be expected that the limit of large $w_z$ corresponds to the fully-supervised model. However, this is not the case; increasing $w_z$ while still using a triple loss resulted in performance worse than the fully-supervised CNN model. 

\section{COSMOS and CANDELS Spectroscopic Redshift Sources} \label{app:spec_z_citations}
This work has greatly benefited from publicly available spectroscopic redshifts. We acknowledge the substantial effort of the teams who released and compiled these datasets. The spectroscopic redshifts we used from \citealt{2023ApJ...942...36K} were compiled from the following works:
\citealt{2003SPIE.4834..161D,
2003SPIE.4841.1657F,
2004A&A...421..913W,
2004AJ....127.3121W,
2004ApJ...601L...5V,
2004ApJ...617..746D,
2004ApJS..155..271S,
2005A&A...437..883M,
2005A&A...439..845L,
2007A&A...465.1099R,
2007ApJ...660L...1D,
2007ApJ...669..776K,
2007ApJS..172...70L,
2008A&A...478...83V,
2008ApJ...689..687B,
2009A&A...494..443P,
2009ApJ...693.1713T,
2009ApJ...695.1163V,
2009ApJ...706..885W,
2009ApJS..184..218L,
2010A&A...512A..12B,
2010ApJS..191..124S,
2011ApJ...741....8C,
2011ApJS..193...14C,
2012MNRAS.425.2116C,
2013A&A...549A..63K,
2013A&A...559A..14L,
2013ApJ...772...48P,
2013ApJS..208....5N,
2013MNRAS.428.1088M,
2013MNRAS.433..194B,
2014ApJ...797...17K,
2015A&A...576A..79L,
2015AJ....149..178M,
2015AJ....150..153W,
2015ApJ...801...97S,
2015ApJ...815...98S,
2015ApJS..218...15K,
2015PASJ...67...82A,
2016ApJS..225...27M,
2019MNRAS.488.4565Z}.

The spectroscopic redshifts we used from \citealt{2025arXiv250300120K} were compiled from the following works:
\citealt{2001MNRAS.328.1039C,
2006ApJ...644..100P,
2007ApJS..172...70L,
2007ApJS..172..383T,
2009ApJ...696.1195T,
2009ApJS..184..218L,
2010ApJ...709..572K,
2010ApJ...722..653F,
2011A&A...529A..72F,
2011ApJ...741....8C,
2011ApJ...742..125G,
2011MNRAS.412.2303B,
2012ApJ...744..149H,
2012ApJ...755...26O,
2012ApJ...761..140C,
2012ApJS..200...13B,
2012MNRAS.426.1782R,
2013ApJ...772..113T,
2014ApJ...791...17M,
2014ApJ...797...17K,
2015A&A...575A..40C,
2015A&A...576A..79L,
2015A&A...583A..72P,
2015ApJ...802...31D,
2015ApJ...806L..35K,
2015ApJ...808..139S,
2015ApJ...808..161O,
2015ApJ...808L..33C,
2015ApJS..218...15K,
2015MNRAS.446.2394B,
2015MNRAS.454.3485Y,
2016A&A...592A..78I,
2016ApJ...817..160L,
2016ApJ...822...42O,
2016ApJ...825....4T,
2016ApJ...828...21N,
2016ApJ...828...56W,
2016ApJS..225...27M,
2016MNRAS.456.1195H,
2016MNRAS.457.1739S,
2016MNRAS.457.1888S,
2017A&A...599A..25P,
2017A&A...599A..95G,
2017A&A...600A.110T,
2017ApJ...834..101T,
2017ApJ...840..101C,
2017ApJ...841..111M,
2017ApJ...842...21M,
2017ApJ...845L..16H,
2018A&A...611A..22S,
2018A&A...618A..85S,
2018A&A...619A.147P,
2018A&A...620A.186P,
2018ApJ...858...77H,
2018ApJ...862L..22C,
2018ApJ...869...27V,
2018ApJS..234...21D,
2018ApJS..237...31L,
2018MNRAS.477.2817S,
2018Natur.553..178S,
2019A&A...631A..87V,
2019ApJ...877...81M,
2019ApJ...880..142S,
2019ApJ...886..124W,
2019ApJ...887...55C,
2019ApJ...887..144J,
2019ApJS..241...10K,
2020A&A...633A.159R,
2020A&A...641A.155V,
2020ApJ...888....4S,
2020ApJ...889...93V,
2020ApJ...890...24V,
2020ApJ...890..171J,
2020ApJ...892....8D,
2020ApJ...903....4N,
2020ApJ...903...47F,
2020ApJ...904..107S,
2020ApJ...904..180O,
2020ApJS..249....3A,
2021A&A...649A..78D,
2021A&A...653A..32D,
2021A&A...654A..80S,
2021A&A...654A.121P,
2021ApJ...913..110C,
2021ApJS..256....9S,
2021ApJS..256...44V,
2021MNRAS.500..358B,
2021MNRAS.501.2659L,
2021MNRAS.505.1382M,
2021NatAs...5..485H,
2022A&A...664A.196E,
2022A&A...665A...3J,
2022ApJ...926...37M,
2022ApJ...926..230N,
2022ApJ...929..159C,
2022ApJ...931..160B,
2022ApJ...934..167H,
2022ApJ...938..109F,
2022ApJS..261...12P,
2022ApJS..263...27H,
2022MNRAS.511.6042E,
2023ApJ...943..177M,
2023ApJ...945L...9I,
2023ApJ...954..103S,
2024A&A...683A.205E,
2024A&A...687A.288G,
2024A&A...690A..55S,
2024A&A...690L..16J,
2024AJ....168...58D,
2024ApJ...963...49K,
2024ApJ...963...51F,
2024ApJ...964...69S,
2024ApJ...966...14Z,
2024ApJ...966...36K,
2024ApJ...970...50C,
2024ApJ...977...51F,
2024MNRAS.527.6591B,
2025A&A...693A.309S,
2025A&A...694A.218B,
2025A&A...696L..14S,
2025ApJ...978...90A,
2025ApJ...980L..29A,
2025ApJ...982...27Z,
2025ApJ...985...61F,
2025ApJ...989...31T,
2025ApJ...989L...7T,
2025ApJ...991...37A,
2025ApJ...994...34K,
2025arXiv250314745D,
2025arXiv250511263N,
2026AJ....171...71R}.

\newpage

\bibliography{references, kodra_references, khostovan_references}{}
\bibliographystyle{aasjournalv7}



\end{document}